\documentstyle[12pt,epsfig]{article}

\newcommand{\beq}{\begin{equation}}
\newcommand{\be}{\begin{equation}}
\newcommand{\ee}{\end{equation}}
\newcommand{\eq}{\end{equation}}
\newcommand{\eeq}{\end{equation}}
\newcommand{\av}[1]{\left<#1\right>}
\newcommand{\dt}[1]{\frac{\delta T_#1}{T}}
\newcommand{\bea}{\begin{eqnarray}}
\newcommand{\eea}{\end{eqnarray}}

\newcommand{\lqcd}{\Lambda_{\rm QCD}}
\newcommand{\tqcd}{\theta_{\rm QCD}}

\newcommand{\Hinf}{H_{\rm inf}}
\newcommand{\Power}{{\cal P}}

\def\centeron#1#2{{\setbox0=\hbox{#1}\setbox1=\hbox{#2}\ifdim
\wd1>\wd0\kern.5\wd1\kern-.5\wd0\fi
\copy0\kern-.5\wd0\kern-.5\wd1\copy1\ifdim\wd0>\wd1
\kern.5\wd0\kern-.5\wd1\fi}}
\def\ltap{\;\centeron{\raise.35ex\hbox{$<$}}{\lower.65ex\hbox{$\sim$}}\;}
\def\gtap{\;\centeron{\raise.35ex\hbox{$>$}}{\lower.65ex\hbox{$\sim$}}\;}
\def\gsim{\mathrel{\gtap}}
\def\lsim{\mathrel{\ltap}}

\def\doeack{\footnote{Work supported by the Department of Energy,
                     contract DE-AC02-76SF00515.}}

\def\Title#1{\begin{center} {\LARGE #1 } \end{center}}
\def\Author#1{\begin{center}{ \sc #1} \end{center}}
\def\Address#1{\begin{center}{ \it #1} \end{center}}

\newenvironment{Abstract}{\begin{quotation} \begin{center}
                       ABSTRACT
     \end{center}\bigskip  }{\end{quotation}}

\def\Acknowledgements{\bigskip  \bigskip \begin{center} \begin{large}
             \bf ACKNOWLEDGEMENTS \end{large}\end{center}}
\def\sfig#1#2#3#4#5#6#7#8#9{
 \begin{figure}[t]
 \centering
 \epsfxsize=#7
 \epsfbox[#3 #4 #5 #6]{#2}
 \hspace*{0in}
 \vspace*{#8}
 \caption{#9}
 \label{#1}
 \end{figure}}
\begin{document}
\begin{titlepage}
{\rightline{\begin{tabular}{r}
SLAC--PUB--10030 \\
SU-ITP-03-19 \\
SCIPP-2003-04 \\
hep-th/0409059 \\
\end{tabular}}}

%\vfill
\vspace{.17in}
\Title{Probing a QCD String Axion with \\
\vspace{.12in} Precision Cosmological
Measurements}
\vspace{.08in}
%\vfill
\Author{Patrick Fox$^a$, Aaron Pierce$^b$\doeack, {\rm and} Scott Thomas$^c$}
\Address{$^a$Santa Cruz Institute for Particle Physics \\
1156 High Street, Santa Cruz, CA 95064 USA \\
\vspace{.07in}
$^b$Stanford Linear Accelerator Center, Stanford University\\
    2575 Sand Hill Road, Menlo Park, CA 94025  USA \\
\vspace{.07in}
$^c$Department of Physics, Stanford University, \\
Stanford, CA 94305-4060 USA}
\vspace{.08in}
\begin{Abstract}
String and M-theory compactifications generically have compact
moduli which can potentially act as the QCD axion.
However, as demonstrated here, such a compact modulus can not
play the role of a QCD axion and solve the strong CP problem
if gravitational waves interpreted as arising from inflation
with Hubble constant $H_{\rm inf} \gsim 10^{13}$ GeV are
observed by the PLANCK polarimetry experiment.
In this case axion fluctuations generated during
inflation would leave a measurable isocurvature and/or non-Gaussian
imprint in the spectrum of primordial temperature fluctuations.
This conclusion is independent of any assumptions about the initial
axion misalignment angle, how much of the dark matter is relic axions,
or possible entropy release by a late decaying particle such as
the saxion; it relies only on the mild assumption that the
Peccei-Quinn symmetry remains unbroken in the early universe.

%String and M-theory compactifications generically have compact
%moduli which can potentially act as the QCD axion and provide an
%elegant solution to the strong CP problem.
%In models in which the gauge couplings naturally unify,
%the Peccei-Quinn scale
%is generally of order the unification scale, $f_a \sim 10^{16}$ GeV.
%Such an axion would likely be impossible to probe directly
%in the laboratory.
%But, as demonstrated here, {\it all} such axion models
%will be definitively excluded for at
%least $f_a \lsim 10^{21}$ GeV
%if primordial gravity waves from an early epoch of inflation
%are observed by the PLANCK polarimetry experiment.
%In this case axion fluctuations generated during
%inflation would leave a measurable isocurvature and/or non-Gaussian
%imprint in the spectrum of primordial temperature fluctuations.
%This conclusion is independent of any assumptions about the initial
%axion misalignment angle, how much of the dark matter is relic axions,
%or possible entropy release by a late decaying particle such as
%the saxion; it relies only on the mild assumption that the
%Peccei-Quinn symmetry remains unbroken in the early universe.

%
\end{Abstract}
\vfill
%\submit{Physical Review {\bf D}}
%\vfill
\end{titlepage}
\tableofcontents
\newpage
\def\thefootnote{\fnsymbol{footnote}}
\setcounter{footnote}{0}
%

% ---------------------------------------------------
% ---------------------------------------------------

\section{Introduction}

M-theory and perhaps its perturbative string theory limits
represent the best hope for a fundamental
theory which integrates matter fields, gauge interactions,
and gravity.
This theory makes many interesting predictions
for physics at the string or eleven-dimensional Planck scales.
These scales could in principle be anywhere
between the electroweak and four-dimensional Planck scales.
However, the only compelling indirect hint for these
scales comes from the unification of gauge couplings
due to four-dimensional renormalization group running below
the unification scale, $M_U \sim 2 \times 10^{16}$ GeV \cite{gcu}.
In this case the fundamental scale is so far above any
energy scale available in the laboratory that direct probes
of string/M-theory seem practically impossible.
However, energies in the early universe might have been
comparable to these scales.
For example, the energy density during inflation could
have been as large as the unification scale.
It is therefore worth investigating in detail whether any
features of string/M-theory could give effects in the
very early universe which might be detected in the
present and coming generation of precision cosmological measurements.

Previous discussions have been rather pessimistic as to the possibility
of discerning effects from such large energy scales
with cosmological measurements \cite{Nemanja}.
However, in this paper we show that there is at least one
feature of many string and M-theory vacua
which can be probed by precision cosmological
measurements -- a string/M-theory QCD axion.
%with Peccei-Quinn scale of order the unification scale.
This conclusion turns out to be nearly independent of
any cosmological assumptions such as the initial axion misalignment
angle, how much of the dark matter is relic axions,
or dilution of relic axions by a late entropy release.

A generic feature of supersymmetric string and M-theory compactifications
is the existence of complex moduli fields.
The compact components of these moduli ultimately
arise in one description or another from $p$-form gauge fields
on non-trivial $p$-cycles in an internal manifold.
The vacuum angles for low energy gauge groups often
transform inhomogeneously under shifts of these compact
moduli.
In the case of QCD such a compact modulus then gives
a realization of the Peccei-Quinn symmetry in Nambu-Goldstone
mode \cite{axions}.
Supersymmetry breaking generically lifts moduli.
While not guaranteed, it is possible that the compact components of moduli
are protected from obtaining a potential from supersymmetry
breaking to a fairly high order by approximate or
exact discrete symmetries \cite{BanksDine}.
In the case of QCD the leading contributions to the potential
for such a compact modulus can then arise from infrared QCD effects.
Such a compact modulus then acts as an invisible axion and provides an
elegant solution to the strong CP problem.

%In particular, since the shift symmetry of the model independent
%axion of perturbative string theory remains unbroken
%at at orders in perturbation theory, it is possible
%that it plays the role of a QCD axion \cite{BanksDine}.

In the classes of vacua mentioned above in which gauge coupling
unification arises from four-dimensional renormalization group
running, the Peccei-Quinn scale for a string/M-theory QCD axion
is generally of order the unification scale, $f_a \sim M_U$.
For example, as presented in appendix A, the Peccei-Quinn scale
for the canonically normalized
model-independent string axion is
\beq
f_a = {\sqrt 2} ~{ \alpha_U \over 4 \pi} M_P \sim 10^{16}~
   {\rm GeV}
\eq
where $\alpha_U = g_U^2 / 4 \pi$ is the unified value of the fine structure
constant and $M_p \simeq 2.4 \times 10^{18}$ GeV is the four-dimensional reduced
Planck mass.
%and the QCD vacuum angle is related to the model-independent
%axion by $\theta_{\rm QCD} = a/f_a$.
We therefore focus on Peccei-Quinn scales
$f_a \sim 10^{16}$ GeV, but consider the general
case of large $f_a$.
The interactions of such axions are exceedingly weak, first because
of the derivative coupling, and second because of the
large Peccei-Quinn scale.
%Probing such an axion in laboratory experiments seems practically
%impossible.
Even if a relic condensate of such an axion
comprised all the dark matter it seems practically impossible
to detect directly in laboratory experiments.

Precision cosmological measurements do, however, provide
the possibility to probe a string/M-theory QCD axion.
Quantum fluctuations are imprinted on such an axion during
inflation.
These fluctuations eventually contribute to temperature fluctuations
in the Cosmic Microwave Background Radiation (CMBR) with the
distinctive characteristics of
isocurvature \cite{AxionIso} and non-Gaussian \cite{lythstewart}
components.
The magnitude of these fluctuations are controlled by the
energy scale of inflation.
At present there is no direct experimental indication for this scale.
%Because there is a great deal of uncertainty in this scale at present, is not
%clear how big of an effect to expect.
However, upcoming experiments sensitive
to the polarization of the CMBR, such
as the PLANCK Surveyor \cite{Planck} or CMBPol could conceivably detect
background primordial gravitational waves.
The presence of a gravitational
wave signal in CMBR fluctuations
would represent a true triumph for CMBR experiments,
and if interpreted as arising from inflation
would imply a rather high inflationary energy scale.
Most importantly for the present discussion,
it would establish the magnitude of primordial axion fluctuations
imprinted during inflation.
And for a given Peccei-Quinn scale, $f_a$, this would
provide a lower limit on the isocurvature and non-Gaussian
components in the CMBR fluctuations.
So if gravity waves are detected in future CMBR fluctuations
experiments, then the existence of a string/M-theory QCD axion
can be tested by a search for isocurvature and/or non-Gaussian
components also in CMBR fluctuations.

%A high scale for inflation
%gives the possibility of large isocurvature perturbations or non-Gaussianities
%arising from the axion, thereby rendering the QCD string axion scenario
%testable.

The notion that information might be gained about axions from the CMBR is not
new 
%\cite{Lyth,LindeAxions,turnerwilczek,lythstewart,Battye,Burns,MTheoryAxions}
[6,8--13].
%There has even been some discussion
%of the possibility of learning about string theory axions in this manner
%\cite{MTheoryAxions}.  
However, these works 
either assume that the axion makes up all of the dark matter 
(or even the entire critical density) 
%\cite{Lyth,LindeAxions,Battye,MTheoryAxions}, 
[8,9,11--13],
do not allow for the possibility of dilution of the axion 
condensate by a late entropy release 
%\cite{Lyth,LindeAxions,turnerwilczek,lythstewart,Battye,Burns,MTheoryAxions},
[6,8--13], and 
do not consider separately isocurvature and non-Gaussian 
contributions to CMBR temperature fluctuations 
[6,8--13].

The goal here is to assess how well precision CMBR measurements
can probe a string/M-theory QCD axion {\it independent}
of any cosmological assumptions.
In order to do this the correlation between the magnitude
of inflationary gravity wave contributions to CMBR fluctuations,
and axion induced isocurvature and non-Gaussian components
of CMBR fluctuations
must be derived for the widest possible range of cosmological assumptions:
First, the initial axion misalignment angle must be taken to be
arbitrary within allowed constraints.
Second, no assumption must be made about how much of the dark matter is
made of relic axions.
And third, any allowed, but otherwise arbitrary, late entropy release which
could dilute any relic axions must be allowed.
Obtaining a definitive statement about how well precision
CMBR measurements can probe a string/M-theory QCD axion then amounts
to employing the cosmological assumptions which
{\it minimize} the axion induced isocurvature and non-Gaussian
contributions to the CMBR fluctuations
for a given magnitude of inflationary gravity wave contributions
to CMBR fluctuations and Peccei-Quinn scale.
This corresponds to assuming a vanishingly small
average misalignment angle, 
with the relic axion condensate arising only from 
inflationary induced fluctuations of the axion 
field \cite{turnerwilczek}, along with 
the maximum allowed dilution of this condensate by a late
entropy release.
In this case, any fluctuation induced relic axions turn out to
make up at most a very small fraction of the dark matter.
If this most pessimistic of cosmological assumptions can
be probed experimentally, then all possible assumptions
can be probed: the observational effects of an axion are
necessarily larger for non-vanishing average initial
misalignment angle and non-maximal dilution by an entropy release.
The motivation for employing
%most
%pessimistic possible assumptions which minimize
%the observational effects of an axion
the widest possible set of cosmological assumptions
is that in the era of
precision cosmological measurements, ideally it should not be necessary
to make {\it ad hoc} untestable assumptions about early universe cosmology
in order to obtain definitive tests.

%We divorce ourselves from
%these assumptions and try to make a critical assessment of what it would mean
%for string QCD axions if gravitational waves were detected at a future
%experiment, independent of assumptions about specific models of inflation.

Even with the above assumptions which minimize the observational effects
of relic axions we come to a strong conclusion:

\vspace{-.1in}
\begin{quote}
{\it A string/{\rm M}-theory compact modulus can not play the role
of a~{\rm QCD} axion and solve the strong~{\rm CP} problem
if gravitational waves interpreted as
arising from inflation  with a Hubble constant
$H_{\rm inf} \gsim 10^{13}$ {\rm GeV} are observed by the {\rm PLANCK}
polarimetry experiment.}
\end{quote}
\vspace{-.1in}

\noindent
This conclusion requires only the current bounds on isocurvature
and non-Gaussian contributions to CMBR fluctuations and
relies on the expectation that the maximum
possible Peccei-Quinn scale for a string/M-theory axion is the
four-dimensional Planck scale
$f_a \lsim M_p$ \cite{Banks:2003sx}.
In terms of an arbitrary Peccei-Quinn scale
these bounds would require
at least $f_a \gsim 3 \times  10^{20}$ GeV
if PLANCK observed gravity waves corresponding
to an inflation scale of $H_{\rm inf} \simeq 10^{13}$ GeV,
and at least $f_a \gsim 5 \times 10^{25}$ GeV if gravity
waves were observed just below the current WMAP bound 
of $H_{\rm inf} \lsim 3 \times 10^{14}$ GeV \cite{wmapinf}.
These bounds do require the mild assumption that the Peccei-Quinn
symmetry remains unbroken in the early universe.
They would be further strengthened by (likely) improvements in the
bound on the isocurvature %and non-Gaussian
components of CMBR fluctuations.

In the next section the production of relic axions is reviewed
in all the relevant cosmological scenarios including the
effects of inflationary induced axion fluctuations and the
possibility of dilution
by an entropy release due a late decaying particle.
The specific case of the late decay of the saxion superpartner
of the axion is considered, but shown to generally lead to
additional cosmological problems rather than a successful
dilution of the axion condensate.
The standard results are also extended to the case of large
Peccei-Quinn scales.
%In Sec.~\ref{sec:relic}, we calculate the minimum axions abundances
%in a variety of cosmological scenarios for formation of the axion
%condensate.
%This includes the possibility of dilution of the axion condensate
%after formation by late decay of a massive particle.
%The minimum inflationary fluctuation
%dominated relic axion abundance which is obtained for very
%small initial misalignment angles is also presented.
In section~\ref{sec:Sigs} the isocurvature and non-Gaussian contributions
to CMBR temperature fluctuations arising from inflationary axion fluctuations
are presented.
The current bounds and possibilities for future improvements on
the CMBR probes of the Hubble constant during inflation
and isocurvature and non-Gaussian components are also discussed.
In section \ref{sec:Combine} the rather stringent bounds 
on the Peccei-Quinn scale which would result
from the current isocurvature and non-Gaussian CMBR temperature
fluctuation bounds are presented in the case that
primordial gravity waves interpreted as arising from inflation
are in fact observed in future experiments.
The implications for a string/M-theory QCD axion are also considered.
In section 5 the scenarios for realizing the Peccei-Quinn mechanism
to which the stringent bounds would apply are spelled out.
Possible loop holes for these bounds are also addressed.
In section 6 the strong conclusion that a string/M-theory QCD
axion is not consistent with the observation of primordial
gravity waves by the PLANCK polarimetry experiment is
reviewed.
And a corollary is presented
regarding the implications of observation
of isocurvature and/or non-Gaussian components of CMBR temperature
fluctuations would have if interpreted as arising from a QCD axion.

The relation between the unified fine structure constant and
properly normalized Peccei-Quinn scale for the model independent
string axion is derived in appendix A.
The parameterization of the finite temperature axion mass
resulting from the dilute instanton calculation of the
$\theta_{\rm QCD}$-dependence of the finite temperature free energy
is given in appendix B.
The relation between the dimensionless skewness defined
in section 3.3 and a parameterization of non-Gaussian
components of temperature fluctuations commonly
used in the literature is derived in appendix C.

% --------------------------------------------
% --------------------------------------------

\section{Relic Axions}
\label{sec:relic}

Relic axions can be produced in the early universe through a variety of
mechanisms: thermal production, axion string radiation, or
condensate formation from misalignment of the initial QCD vacuum
energy \cite{oldrelic,turner,turnerwindows,KawasakiMoroi}.
For the large Peccei-Quinn scales relevant here, thermal production
is unimportant.
In addition, since any string/M-theory
modulus which contains a QCD axion is present before and during
inflation any axion strings are inflated away.
This leaves coherent production from misalignment as the
dominant mechanism for producing relic string/M-theory QCD axions.

The formation of a QCD axion condensate is reviewed in
section 2.1, with special attention paid to the case of
large Peccei-Quinn scale relevant to a string/M-theory axion.
It is shown that there are sizeable uncertainties in the
calculation of the axion relic density due to strong QCD effects
for Peccei-Quinn scales
%at least in the range
%$10^{15}~{\rm GeV} \lsim f_a/N \lsim 10^{17}~{\rm GeV}$.
in the region of interest, but that this does not affect
the final conclusions.
An updated version of the classic cosmological bound on
the Peccei-Quinn
scale for an average misalignment angle is presented
taking into account recent determinations of the dark matter density.
This bound is significantly exceeded for $f_a/N \sim 10^{16}$ GeV,
the value that might be expected for a string/M-theory QCD axion.
So the possibility of accommodating such an axion by reducing
relic production
with a small misalignment angle is considered in section 2.2.
The complimentary or additional possibility of reducing relic
axion density by a late entropy release is discussed in section
2.3.
It is shown that the maximum possible dilution by a late
decaying particle is alone not sufficient to accommodate
$f_a/N \sim 10^{16}$ GeV with average misalignment angle.
Such an axion requires a small misalignment angle.
The inflationary quantum fluctuations imprinted on an
axion are reviewed in section 2.4.
These fluctuations are shown to lead to a minimum relic
density of axions which depends on the Hubble constant
during inflation and the Peccei-Quinn scale.

%For the large axion decay constant relevant for a QCD string axion the
%calculation of the axion relic density in a radiation dominated
%universe is subject to large uncertainties, due to strong QCD effects.
%The calculation for the case of a matter dominated universe does not
%suffer from these uncertainties.

%\cite{turner,turnerwindows,KawasakiMoroi},
%\cite{Bunch}

% ---------------------------------------------------

\subsection{Formation of the Axion Condensate}

A relic axion condensate is formed
in the early
universe from misalignment of the initial QCD vacuum angle \cite{oldrelic}
when the axion begins to
oscillate at $3H \simeq m_a(T_{\rm osc})$, where $H$ is the Hubble parameter.
Given the temperature dependence of the axion mass $m_a(T)$,
this condition along with the Friedman equation,
$3 H^2 M_p^2 = \rho$, and thermal energy density
in a radiation dominated era
$\rho = (\pi^2/30) g_{*} T_{\rm osc}^4$ defines the
temperature, $T_{\rm osc}$,
at which the axion begins to oscillate
\beq
T_{\rm osc} \simeq \bigl( 10/ g_* \pi^2 \bigr)^{1/4}
    \bigl( m_a(T_{\rm osc}) M_p \bigr)^{1/2}
\label{Tosceq}
\eq
For oscillation temperatures just above $\Lambda_{\rm QCD}$,
$g_{*} \simeq 61.75$.
The parameterization of the high
temperature dependence of the axion mass for $T \gsim \Lambda_{\rm QCD}$
is reviewed in Appendix \ref{appmass}, and may be written \cite{gpy,turner}
\beq
\xi(T) \equiv{m_a(T) \over m_a }  \simeq
  C \left( \Lambda_{\rm QCD} \over 200 ~{\rm MeV} \right)^{1/2}
  \left( \Lambda_{\rm QCD} \over T \right)^{4}
  ~~~~~~{\rm for}~~T \gsim \Lambda_{\rm QCD}
%  \left[ 1 - \ln\left( \Lambda_{\rm QCD} \over T \right)
%    \right]^d
\label{fitxi}
\eeq
where $C\simeq 0.018$ is a parameter defined in Appendix \ref{appmass}.
With this, the oscillation temperature
%for $T_{\rm osc} \gsim \Lambda_{\rm QCD}$
is then
\beq
T_{\rm osc} \sim %260~{\rm MeV} ~
  150~{\rm MeV} ~
 \left( { C \over 0.018} \right)^{1/6}
 \left( { \lqcd \over 200~{\rm MeV} } \right)^{3/4}
 \left( { 10^{16}~ {\rm GeV} \over f_a / N} \right)^{1/6}
   ~~~~{\rm for}~~  T_{\rm osc} \gsim \Lambda_{\rm QCD}
\label{Tosc}
\eq
This expression for the oscillation temperature is only appropriate
for $T_{\rm osc} \gsim \Lambda_{\rm QCD}$ corresponding
to $f_a/N \lsim 2\times 10^{15}$ GeV.
The case of $T_{\rm osc} \lsim \Lambda_{\rm QCD}$ for larger
Peccei-Quinn scales is considered separately below.

The local axion number density in the condensate when it is formed
is
\beq
n_a \simeq f_c {1 \over 2}
   %m_a(T_{\rm osc})
   \xi(T_{\rm osc}) m_a
   (f_a/N)^2 (\theta_i^2+ \sigma_{\theta}^{2}) f(\theta_i^2)
\label{relicna}
\eq
where $m_a=m_a(0)$ is the zero temperature mass,
$f_c$ is a correction for temperature dependence of the
axion mass, $f_c \simeq 1.44$ \cite{turner}
for the $\xi(T)$ parameterization (\ref{fitxi}),
and $f(x)$ is a correction for anharmonic effects of the
axion potential; for $x \ll 1$, $f(x) \rightarrow 1$.
And $\theta_i \equiv \langle \theta \rangle$ is the
initial QCD vacuum angle zero-mode averaged over the presently
observable universe,
and $\sigma_{\theta}^2 \equiv \langle ( \theta - \langle \theta
\rangle)^2 \rangle$ is the mean square fluctuations of the initial
QCD vacuum angle, produced for example during an early epoch
of inflation.
Since the number density redshifts like the entropy density,
the local axion density today is given by
\beq
\rho_a \simeq (n_a/s)m_a s_0 \gamma
\label{relicrhoa}
\eq
 where $s=(2 \pi^2 / 45) g_{*s} T_{\rm osc}^3$ is the thermal
entropy density when the axion begins to oscillate,
$ s_0 \simeq (2 \pi^2 / 45) g_{*s0} T_{0}^3 $ is the thermal
entropy density today with $g_{*s0} \simeq 3.91$ and
$T_0 \simeq 2.73$ K.
The factor $\gamma$
is a possible dilution factor
due to entropy release after the axion begins to oscillate,
$\gamma =  S_{\rm osc}/S_0$.
Using all this with the oscillation temperature (\ref{Tosc})
leads to a relic axion density,
$\Omega_a = \rho_a / (3 H_0^2 M_p^2)$, of
%\begin{eqnarray}
%\label{relicdensity}
%\Omega_a h^2 & \sim &% 2 \times 10^4
%  (2 \times 10^4)   f(\theta_i^2) (\theta_i^2 + \sigma_{\theta}^2)) \gamma \\
% &\times& \left( { 0.018 \over C } \right)^{1/6}
% \left( { 200~{\rm MeV} \over \lqcd } \right)^{3/4}
% \left( { f_a / N \over 10^{16}~ {\rm GeV}} \right)^{7/6}
% \left( {g_{*}(T_{osc}) \over 61.75} \right)^{3/4}  \nonumber
%\end{eqnarray}
\beq
\Omega_a h^2  \sim  2 \times 10^4
 \left( { 0.018 \over C } \right)^{1/6}
 \left( { 200~{\rm MeV} \over \lqcd } \right)^{3/4}
 \left( { f_a / N \over 10^{16}~ {\rm GeV}} \right)^{7/6}
 \left( \theta_i^2 + \sigma_{\theta}^2 \right)
 f(\theta_i^2) \gamma
\label{relicdensity}
\eq
where $h=H_0/(100~{\rm km}~{\rm s}^{-1}~{\rm Mpc}^{-1})$.
This expression for the relic axion density is only valid
for at least $f_a/N \lsim 2 \times 10^{15}{\rm GeV}$.

Since the expression (\ref{relicdensity}) for the relic axion density
uses the high temperature expression (\ref{Tosc})
for the oscillation temperature,
it does not apply for $T_{\rm osc} \lsim \Lambda_{\rm QCD}$.
For oscillation temperatures in this range (corresponding to
large Peccei-Quinn scales), the axion begins to oscillate
in effectively the zero temperature potential and
another expression for the relic density is obtained.
For oscillation temperatures just below $\Lambda_{\rm QCD}$,
$g_{*} \simeq 10.75$.
With this, and for a general temperature dependence for
the axion mass, $\xi(T)$, the oscillation temperature
(\ref{Tosceq})
%for $T_{\rm osc} \lsim \Lambda_{\rm QCD}$
is
\beq
T_{\rm osc} \sim 950~{\rm MeV}~ \left( { 10^{16}~ {\rm GeV} \over f_a}
   \right)^{1/2} ~ \xi(T_{\rm osc})^{1/2}
   ~~~~~{\rm for}~~T_{\rm osc} \lsim \Lambda_{\rm QCD}
\label{eqn:Tosczero}
\ee
This expression %(\ref{eqn:Tosczero})
for the oscillation temperature is only appropriate
for $T_{\rm osc} \lsim \Lambda_{\rm QCD}$, corresponding to
at least $f_a/N \gsim 2 \times 10^{17} ~\xi(T_{\rm osc})$ GeV.
For $T_{\rm osc} \lsim \lqcd$, the relic axion density from
(\ref{relicna}) and (\ref{relicrhoa}) is
\beq
\Omega_a h^2 \simeq 5 \times 10^3
 \left( { f_a / N \over 10^{16}~ {\rm GeV}} \right)^{3/2}
 \left( \theta_i^2 + \sigma_{\theta}^2 \right)
 f(\theta_i^2) %(\theta_i^2 + \sigma_{\theta}^2))
 \gamma ~f_c \xi(T_{\rm osc})^{-1}
 \label{relicdensitylargef}
\eq
While there is some residual temperature dependence of the axion mass
for $T_{\rm osc} \lsim \Lambda_{\rm QCD}$, this is unimportant and
$f_c \xi(T_{\rm osc})^{-1} \simeq 1$ is a good approximation well into
this regime.
The relic density
(\ref{relicdensitylargef}) is only valid for
at least $f_a/N \gsim 2 \times 10^{17}$ GeV.
Note that is has a different parametric dependence on
$f_a/N$ than (\ref{relicdensity}) appropriate for small
Peccei-Quinn scales.

Note that the relic axion densities (\ref{relicdensity}) and
(\ref{relicdensitylargef})
calculated for oscillations temperatures
(\ref{Tosc}) and (\ref{eqn:Tosczero}) above and below
$\lqcd$ do not have overlapping regions of
validity.
This indicates that in the transition region, $2\times
10^{15}~{\rm GeV}\lsim f_a/N\lsim 2\times 10^{17}~{\rm GeV}$,
neither expression is completely accurate.
This is in fact to be expected for $T_{\rm osc} \sim \lqcd$.
The high temperature expression (\ref{Tosc}) is derived using a
dilute instanton
gas approximation \cite{gpy} as described in Appendix \ref{appmass}.
This likely receives non-trivial corrections from
multi-instantons for $T_{\rm osc}$ somewhat above $\lqcd$.
Likewise the low temperature expression (\ref{eqn:Tosczero})
likely receives non-trivial corrections from
spontaneous chiral symmetry breaking effects for
$T_{\rm osc}$ just slightly below $\lqcd$.
It is perhaps unfortunate that for a Peccei-Quinn
scale relevant to a string/M-theory axion, strong QCD effects are
important during formation of the axion condensate.
However, for $f_a/N$ well outside the transition region
the true expression for the axion relic density asymptotes to
the expressions given above, and should smoothly interpolate
between these limiting forms.
And the bounds described below turn out to apply to all
Peccei-Quinn scales up to a very large value which is well
outside the transition region.
So the final conclusions are not sensitive to unknown
strong QCD effects.
The relic densities (\ref{relicdensity}) and (\ref{relicdensitylargef})
turn out to be equal
for $f_a/N \simeq 6 \times 10^{17}$ GeV.
This is in fact outside the transition region given above, due
mainly to the abrupt change in the number of degrees
of freedom at a temperature $T \sim \lqcd$.
In what follows
we simply use (\ref{relicdensity}) for Peccei-Quinn scales lower than
this value and (\ref{relicdensitylargef}) for higher scales.
%with the
%understanding that there are large uncertainties in the transition
%region, as discussed above.
Also note that the expressions for the
relic densities (\ref{relicdensity}) and
(\ref{relicdensitylargef}) only apply if the universe is radiation
dominated when the axion condensate forms.
The case of matter domination is presented in
section~\ref{sec:matterdom}.

%Below we simply use (\ref{Tosc}) for $f_a/N \lsim~{\rm few}~\times
%10^{16}$ GeV with the understaning that
%there are unknown uncertainties when this limit is nearly
%saturated.
%%The oscillation temperature (\ref{Tosc})

Most of the relic axions have momenta much less than the
zero temperature axion mass, $p_a \ll m_a$, and so
act as cold dark matter (CDM).
The present CDM energy density has been
accurately determined in a $\Lambda$CDM cosmology
by recent measurements
to be $\Omega_{\rm CDM} h^2=
0.113 \pm 0.010 $ \cite{WMAP}.
At most, relic axions could saturate this density and make
up all the CDM.
Alternately, the CDM might consist of more than
one species, with relic axions contributing only part of the
total observed CDM density.
So the relic axion density is bounded from above by
\beq
\Omega_a h^2 \lsim 0.12
\label{omegabound}
\eq

The average mean square initial misalignment angle over all
inflationary domains is
$\langle \theta^2 \rangle = \pi^2 / 3$, for which
the anharmonic correction is $f(\pi^2/3) \simeq 1.2$ \cite{turner}.
So for the average misalignment angle the
relic axion density (\ref{relicdensity})
without any dilution, $\gamma=1$, along with the
bound (\ref{omegabound}), require that the Peccei-Quinn scale
is bounded from above by
\beq
f_a/N \lsim 3 \times 10^{11}~{\rm  GeV}
~~~~~{\rm for}~~\langle \theta^2 \rangle = \pi^2 / 3~~
{\rm and}~~\gamma=1
\label{newbound}
\eq
This is somewhat smaller than the canonical $f_a/N \lsim 10^{12}$ GeV
bound arising from $\Omega_a h^2 \lsim 1$ quoted in the older
literature \cite{oldrelic} mainly due to the recent improvement
in measurement of the CDM density quoted above.

% ---------------------------------

\subsection{Small Initial Misalignment Angle}

A QCD string/M-theory axion with $f_a/N \sim 10^{16}$ GeV is
clearly inconsistent with the bound (\ref{newbound}) for
the average misalignment angle
and without any dilution of the axion condensate after its formation.
One possibility which could accommodate such an axion
is that the average misalignment angle within
the observable universe is small.
For the relic axion density (\ref{relicdensity})
with $\gamma=1$, the bound (\ref{omegabound}) then requires
\beq
(\theta_i^2 + \sigma_{\theta}^2)^{1/2} \lsim 2 \times 10^{-3}~
\left({10^{16}~{\rm GeV} \over  f_a/N} \right)^{7/12}
\label{thetaibound}
\eeq
In fact, with random initial conditions prior to an early
epoch of inflation, there are at present distant regions of the universe
which sample all values of initial misalignment angles.
So it is possible in principle
that $\theta_i^2 + \sigma_{\theta}^2$ happens to be sufficiently
small in our observable universe to allow $f_a/N \sim 10^{16}$ GeV.
Regions with misalignment angles which are
larger(smaller) than ours then have
a larger(smaller) Hubble constant at a given temperature.

In this scenario the question naturally arises though
as to why our initial misalignment angle is so small.
Linde has argued that
with $f_a$ as large as considered here there may be a dynamical
selection effect which makes the observation of small misalignment
angle likely \cite{LindeAxions}
(for a recent discussion see \cite{wilczek}).
At fixed large $f_a$, the density of gravitationally collapsed objects
such as galaxies grows as a rapid power of $\theta_i^2$
\cite{LindeAxions}.  Since very dense galaxies may be inhospitable to
observers such as ourselves, it may then be more likely that a given
observer measures a small misalignment angle.  However, quantifying
the probability distribution of observers as a function of galactic
density remains an open problem.
In addition, such arguments could not explain an initial
misalignment angle significantly smaller than the bound
(\ref{thetaibound}).

Independent of the question of
possible selection effects for the observed value of the
initial misalignment angle,
it is possible
in principle that the misalignment angle in the observable universe
is so small for whatever reason
that relic axions comprise a vanishingly small fraction of the CDM.
In this case it might appear that there are no observational
effects.
However, quantum fluctuations of the axion during inflation
lead to a non-vanishing mean square fluctuation of the QCD
vacuum angle, $\sigma_{\theta} \neq 0$, as discussed in
section \ref{infaxionsection}.
This gives a lower limit on the relic axion density
(\ref{relicdensity}) or (\ref{relicdensitylargef}),
and therefore on the observable effects.
The magnitude of these effects depends crucially on the
Hubble constant during inflation, as discussed below.
The goal here is to establish how well a QCD string/M-theory
axion can be probed by precision cosmological measurements
independent of any assumptions about the initial misalignment
angle.
To do this, the Hubble constant during inflation must therefore be measured
by independent means in order to establish a minimum relic axion density,
as discussed in section \ref{gwaves}.

% --------------------------------

\subsection{Dilution of the Axion Condensate}
\label{sec:matterdom}

Another possibility which might help accommodate a QCD string/M-theory
axion is that there is a late entropy release
with a reheat temperature
$T_{\rm RH} < T_{\rm osc}$ which dilutes the axion
condensate after its formation \cite{dilutionref}.
For the relic axion density (\ref{relicdensity}) with
average misalignment angle
$\langle \theta^2 \rangle = \pi^2 / 3$,
the relic density bound (\ref{omegabound})
requires a dilution factor of
\beq
\gamma \lsim 2 \times 10^{-6}~
\left( { 10^{16}~{\rm GeV} \over f_a/N } \right)^{7/6}
\eeq

A late entropy release must satisfy a number of non-trivial
constraints in order to successfully dilute the axion condensate.
First, the entropy release must occur after the condensate is
formed at $T_{\rm osc} \sim \Lambda_{\rm QCD}$.
An entropy release at a higher temperature does not modify
the relic axion density.
Second, the entropy release must not adversely modify the
successful predictions of relic light element abundances
from big bang nucleosynthesis (BBN).
The $^4{\rm He}$ abundance is determined in standard
radiation dominated BBN mainly by the primordial baryon to
entropy ratio, $n_b/s$, and the neutron to proton ration, $n/p$,
when the weak interactions drop out of equilibrium at $T \sim 1$ MeV.
Standard BBN then generally requires that any entropy release terminate
in a radiation domination era with a reheat temperature of
at least $T_{\rm RH} \gsim 1$ MeV and possibly higher.
Finally, the entropy
release must not over-dilute the primordial relic baryon density.

The case of an entropy release by a late decaying particle
is considered below.
The special case of a late decaying saxion superpartner of the axion
is also presented.
The very unlikely possibilities of a first order phase
transition or very late inflation which
could dilute the axion condensate are commented on in
section \ref{windows}.

% ------------------------------------------------------

\subsubsection{Late Decaying Particle}

A very natural way to obtain a late entropy release is
by the late decay of a massive non-relativistic particle.
In order to have a non-trivial dilution of the axion
condensate the late decaying particle must dominate
the energy density for at least
$T_{\rm osc} \gsim \Lambda_{\rm QCD}$.
This can occur in a number of ways.
For a scalar particle, a relic condensate formed from an
initial misalignment of the scalar field from the minimum
of the potential can dominate at a very early epoch.
For either a scalar or fermion particle, relic particles
produced during reheating after inflation, or simply from
thermal freeze out can dominate at an early epoch.
For the discussion here, detailed models for the origin of
the late decaying relic particles are not necessary.
The dilution of the axion condensate can be parameterized
solely in terms of the particle reheat temperature as
described below.

For a late decaying particle the BBN bounds on the reheat
temperature are slightly more stringent than the
case of a general entropy release discussed above.
An exponentially small number of late decaying particles
survive to temperatures below the reheat temperature.
If hadronic decay channels are available (which is
inevitable in the thermalization process)
then these very late decays can enrich $n/p$ even for
$T_{\rm RH} \gsim 1$ MeV.
And this can be a significant effect since $n_b/s$ is so small.
For a late decaying particle with $m \sim 1$ TeV, a reheat
temperature of $T_{\rm RH} \gsim 6$ MeV is required in order
to ensure that the $^4{\rm He}$ abundance is not too
large \cite{nucleomev}.
So in order to dilute the axion condensate while keeping
the successful predictions of BBN a late decaying
particle must first, dominate the energy density for at least
$T \gsim \Lambda_{\rm QCD}$ and second, have a reheat temperature
in the range $6~{\rm MeV} \lsim T_{\rm RH} \lsim \Lambda_{\rm QCD}$.

The relic axion density which results with the late decay of a massive
particle $\phi$ can be calculated
from the axion fractional energy
density when the axion condensate is formed.
If a massive non-relativistic particle dominates the energy
density, the universe is matter dominated
at this epoch with $\rho_{\phi} =n_{\phi} m_{\phi} \simeq 3 H^2 M_p^2$.
The number density in the axion condensate when it forms
at $3H \simeq m_a(T_{\rm osc})$ is given by (\ref{relicna}).
Since the axion and massive particle number densities
both redshift like matter, the ratio $n_a/n_{\phi}$ is constant
after the axion condensate is formed and before the massive
particle decays.
Using all this, the ratio of axion density $\rho_a = m_a n_a$
to massive particle density during this epoch is
\beq
{\rho_a \over \rho_{\phi}} \simeq f_c {3 \over 2} \xi(T_{\rm osc})^{-1}
 { (f_a / N)^2 \over M_p^2}  \left( \theta_i^2 + \sigma_{\theta}^2 \right)
f(\theta^2_i)
\label{ratioaphi}
\eq
When the massive particle decays its energy is converted
to thermal radiation.
In the sudden decay approximation the resulting thermal
energy and entropy densities are
$ \rho_{\phi} \simeq (\pi^2 / 30) g_* T_{\rm RH}^4 $
and
$s \simeq (2 \pi^2 / 45) g_{*} T_{\rm RH}^3$.
The relic axion number density to entropy density ratio at this epoch is then
\beq
{n_a \over s} = {\rho_a /m_a \over \rho_{\phi} }
  {\rho_{\phi} \over s}
  \label{asdecayrat}
\eq
with $\rho_{\phi}/s \simeq (3/4) T_{\rm RH}$.
Since the number density redshifts like the entropy density
after the late decay,
the local axion density today is given by (\ref{relicrhoa}) using
the ratio (\ref{asdecayrat}).
Combining all this, the relic axion density with the
late decay of a massive particle is \cite{KawasakiMoroi}
\beq
\Omega_a h^2 \simeq
  35 \left( { T_{\rm RH} \over 6~{\rm MeV} } \right)
 \left( { f_a / N \over 10^{16}~ {\rm GeV}} \right)^{2}
 \left( \theta_i^2 + \sigma_{\theta}^2 \right)
 f(\theta_i^2) ~  ~f_c \xi(T_{\rm osc})^{-1}
 \label{rhbound}
\eq
Note that in this scenario appropriate to a matter dominated
era, the relic axion density has a parametric
dependence on $f_a/N$ which is different than both
(\ref{relicdensity}) and (\ref{relicdensitylargef}) appropriate
to a radiation dominated era.

Starting from a very early radiation dominated era,
the temperature for a given Hubble parameter
during the subsequent late decaying particle matter dominated
era is smaller than that for the same Hubble parameter
of a radiation dominated era.
So as long as the late decaying particle dominates the energy
density at a temperature at least somewhat above $\lqcd$, the temperature
that the axion begins to oscillate, $T_{\rm osc}$,
is at least somewhat below $\lqcd$.
The axion condensate then forms in the zero temperature
potential, and $f_c \xi(T_{\rm osc})^{-1} \simeq 1$ is a good approximation.
This implies that if the late decaying particle comes to dominate
the energy density for a temperature well above $\lqcd$
(which is likely in most scenarios) then
the relic abundance (\ref{rhbound}) does not suffer unknown uncertainties
from strong QCD effects in the region of interest,
unlike the relic densities
(\ref{relicdensity}) and (\ref{relicdensitylargef}) appropriate
to a radiation dominated era.

For the average mean square misalignment angle of
$\langle \theta^2 \rangle = \pi^2 / 3$,
the relic density (\ref{rhbound}) with maximum
dilution, $T_{\rm RH} \simeq 6$ MeV,
along with bound (\ref{omegabound}), requires that the Peccei-Quinn scale
is bounded from above by
\beq
f_a/N \lsim 3 \times 10^{14}~{\rm  GeV}
~~~~~{\rm for}~~\langle \theta^2 \rangle = \pi^2 / 3~~
{\rm and}~~
%T_{\rm RH} \simeq 6~{\rm MeV}
{\rm Maximum~Dilution}
\label{farhbound}
\eq
A string/M-theory QCD axion with $f_a/N \sim 10^{16}$ GeV
is inconsistent even with this maximum dilution bound.
So dilution from late decay of a massive particle
turns out alone to be insufficient to accommodate such
an axion with generic misalignment angle.
A small misalignment angle is therefore required in any cosmological
scenario with a Peccei-Quinn scale which far exceeds the bound
(\ref{farhbound}).

% --------------------------------

\subsubsection{Saxion Decay}

In any supersymmetric model the pseudoscalar axion has a
scalar saxion superpartner.
For the model independent string axion the saxion is just
the dilaton.
%
%At least some dilution
%may in fact be likely with a QCD string/M-theory axion.
%In this case the axion is a member of a supermultiplet which
%also contains a bosonic saxion and fermionic axino superpartners.
In the supersymmetric limit, Peccei-Quinn symmetry ensures
that the entire axion supermultiplet remains massless
to all orders in perturbation theory.
However, with supersymmetry and $U(1)_R$ breaking,
only the axion mass is protected by Peccei-Quinn symmetry,
and the saxion potential %and axino mass are
is determined by supersymmetry breaking.
The saxion mass is then naturally much larger than the axion mass.
A coherent condensate of the
saxion can form at a very early epoch from saxion misalignment
and dominate the energy
well before the axion condensate forms.
The saxion condensate eventually decays, with the decay products
rethermalizing to a standard radiation dominated era.
So a supersymmetric axion model naturally contains
all the elements which could in principle dilute the axion
condensate \cite{saxiondecay,SUSYAxions}.
However, as shown below, the reheat temperature is generally
too low for $f_a/N \sim 10^{16}$ GeV.
This complicates rather than improves
the cosmological scenario.

%The saxion begins to oscillate at $3H \simeq m_s$ forming a condensate
%with density $\rho_s \simeq {1 \over 2} m_s^2 s^2$.
%If the saxion misalignment is of order the
%Peccei-Quinn scale, $s \sim f_a$,
%this can be a sizeable fraction of the background energy density
%driving expansion, $\rho = 3 H^2 M_p^2$,
%if $f_a$ is not too much smaller than the Planck scale.
%%If the saxion condensate is formed in a radiation dominated era
%%it can then eventually dominate the energy density since it
%%redshifts like matter.
%The saxion condensate redshifts like matter and
%%This condensate
%can eventually dominate the energy density
%leading to a saxion matter dominated era.

%The condensate ultimately decays, with the decay products
%rethermalizing to a standard radiation dominated era.

The detailed cosmological scenario which is obtained with a
supersymmetric axion model depends on the saxion reheat
temperature.
This is determined by the saxion decay width,
$\Gamma$, and mass,
and is therefore sensitive to details of supersymmetry breaking.
In the sudden decay approximation the saxion condensate decays when
$\Gamma \simeq H$ with its energy converted to thermal radiation,
$( \pi^2 / 30) g_* T_{RH}^4 \simeq 3 \Gamma^2 M_p^2$.
%This reheating can in fact take place after the
%axion condensate forms.
%For example, for a Peccei-Quinn scale and saxion mass of
%$f_a/N \sim 10^{16}$ GeV and $m_s \sim 1$ TeV, depending on the saxion
%decay channels, the saxion reheat temperature can be below the axion
%oscillation temperature.
%$T_{\rm RH} < T{\rm osc}$.
The decay width depends on the saxion couplings which are
model dependent.
However, if the anomalous axion--gluon coupling
(\ref{gaugelag}) given in Appendix A arises directly in the
high energy theory (rather than from integrating out low
energy standard model fields) as is the case for large
classes of string/M-theory axions including
the model independent string axion,
the saxion--gluon coupling is determined by supersymmetry
\beq
{s \over f_a / N} {g_s^2 \over 32 \pi^2} F_{\mu \nu}^a F^{a \mu \nu}
\label{sggcoupling}
\eq
where the gluon fields are canonically normalized.
This coupling gives a decay rate for $s \rightarrow gg$ of
\beq
\Gamma(s \rightarrow gg) = {\alpha_s^2 \over 64 \pi^3 }
  {m_s^3 \over (f_a/N)^2}
\label{saxiongamma}
\eq
In the sudden decay approximation
%the saxion condensate decays when
%$\Gamma \simeq H$ with its energy converted to thermal radiation,
%$( \pi^3 / 30) g_* T_{RH}^4 \simeq 3 \Gamma^2 M_p^2$.
%In this approximation,
the decay rate (\ref{saxiongamma}) gives a reheat
temperature of
\beq
T_{\rm RH} \simeq 1 \times 10^{-2} ~{\rm MeV}~
  \left( { m_s \over {\rm TeV}} \right)^{(n+1)/2}
  \left( { 10^{16}~{\rm GeV} \over f_a/N } \right)
%  ~~~~~~{\rm for}~~~s \rightarrow gg
\label{saxionrh}
\eq
with $n=2$ corresponding to the two derivatives in the
coupling (\ref{saxiongamma}).
As another example, the DFS \cite{dfs} class of axion models
couple only to the visible sector through the Higgs sector.
The couplings to standard model fields then come from
mixings with Higgs bosons.
In this case the leading coupling of the saxion are to heavy quarks
\beq
c ~{ s \over f_a/N } ~m_Q \overline{Q} Q
\label{dfscoupling}
\eq
For the DFS axion $c={1 \over 6}$ for each quark and $N=6$.
Couplings to gluons include an additional loop factor and
are therefore suppressed.
%For a saxion heavier than the top quark $m_s \gsim m_t$, the decay
The coupling (\ref{dfscoupling}) gives a decay rate for
$s \rightarrow \overline{Q}Q$ of
\beq
\Gamma(s \rightarrow \overline{Q} Q ) =
  {3 c^2 \over 16 \pi }
  {m_s m_Q^2 \over (f_a/N)^2}
  \left( 1 - {4 m_Q^2 \over m_s^2} \right)^{3/2}
\label{DFSsaxiongamma}
\eq
For a DFS saxion heavier than the top quark, the dominant decay is
to tops, $s \to t\bar{t}$, with a reheat temperature
from (\ref{DFSsaxiongamma}) which numerically
turns out to be essentially identical
to (\ref{saxionrh}) with $n=0$ corresponding to the absence of derivatives
in the coupling (\ref{dfscoupling}).

For a saxion mass of $m_s \sim 1$ TeV, the reheat temperature
(\ref{saxionrh}) is consistent with the standard
BBN requirement of $T_{\rm RH} \gsim 6$ MeV \cite{nucleomev}
only for at least $f_a/N \lsim 2 \times 10^{13}$ GeV.
So for $f_a/N \sim 10^{16}$ GeV appropriate to a string/M-theory QCD
axion, the rather late decay of a
saxion condensate formed at an early epoch
leads to an additional saxion problem which must be solved, rather
than contributing successfully to dilution of the axion condensate.
A very large saxion mass could however increase the reheat
temperature sufficiently to be consistent with BBN.
For example, in the case of an saxion with high energy
coupling (\ref{sggcoupling}) and Peccei-Quinn scale
$f_a/N \sim 10^{16}$ GeV, a saxion mass of $m_s \gsim 70$ TeV
would just barely give $T_{\rm RH} \gsim 6$ MeV.
Although as discussed in section 2.3.1, dilution by late decay
alone is not sufficient to accommodate such an axion
without small initial misalignment angle.
And obtaining such a heavy saxion seems difficult in most
models of supersymmetry breaking with superpartners just
above the electroweak scale.
In many models of supersymmetry breaking the saxion mass is in
fact much lighter than assumed above, $m_s \ll $ TeV,
leading to an even more severe saxion problem.
In addition, the axino fermionic partner of the axion has, 
for a given mass, a similar
decay rate and reheat temperature to those given 
above \cite{axinoref}.
%But is some scenarios for breaking supersymmetry the axino 
%is even lighter than the saxion. 
So if sufficiently produced, relic axinos can also lead to a problem.
However, in some scenarios for supersymmetry breaking the axino
can be the lighest supersymmetric particle, and might be 
a CDM candidate \cite{axinoref}. 

Any complete cosmological scenario which includes a supersymmetric
QCD axion must address the potential problems associated with
the saxion and axino partners,
presumably by limiting the initial cosmological production of these states.
Since the goal here is to assess how well precision cosmological
measurements can probe a string/M-theory QCD axion
independent of specific cosmological assumptions,
detailed models which address the saxion and axino problems
will not be presented.
Instead, we show below that in any model in which these problems
are in fact solved, precision CMBR measurements can
in principle definitively probe such an axion.

%\beq
%T_{\rm RH} \simeq 8 \times 10^{-3} ~{\rm MeV}~
%  \left( { m_s \over {\rm TeV}} \right)^{1/2}
%  \left( { 10^{16}~{\rm GeV} \over f_a/N } \right)
%  ~~~~~~{\rm for}~~~s \rightarrow \overline{t} t
%\eq

% --------------------------------------------------------------

\subsection{Inflationary Axion Fluctuations}
\label{infaxionsection}

The relic axion density is proportional to the
quadrature sum of the zero-mode
and root mean square fluctuations of the initial QCD vacuum angle,
$\theta_i^2 + \sigma_{\theta}^2$.
This combination amounts to an arbitrary initial condition for the
mean square axion field when the axion condensate is formed.
But in an inflationary cosmology the mean square axion fluctuations
are related to the Hubble constant during inflation.
In this section, inflationary production of axion fluctuations are reviewed.
The minimum relic axion densities associated with these
fluctuations \cite{turnerwilczek}
for the scenarios described in the previous sections are presented.
These minimum relic axion densities are crucial in determining
a lower limit on the observational effects described in the subsequent
sections.

All massless fields undergo de Sitter fluctuations during an
inflationary epoch.
At zero temperature the leading Peccei-Quinn breaking effects
which lift the axion potential must be due
to QCD instantons in order to obtain a successful
solution of the strong CP problem.
These effects are exponentially suppressed during inflation;
and as long as the Peccei-Quinn symmetry is not explicitly
broken in the early universe, the axion is effectively
massless during inflation.
The de Sitter power spectral density of fluctuations of a massless field
such as the axion
\beq
{\cal P}_{a}(k) \equiv { k^3 \over 2 \pi^2} \langle | \delta a_{\bf k} |^2 \rangle
\label{apowerdensity}
\eq
is related to the Hubble constant during inflation by
\beq
{\cal P}_a = \left( { \Hinf \over 2 \pi } \right)^2
\label{poweraH}
\eq
where the ${\bf k}$-space density is $Vd^3k/(2 \pi)^3 =V dk k^2 / 2 \pi^2$
and $V$ is the quantization volume.
The power spectral density (\ref{apowerdensity}) is nothing other than
the mean square fluctuations of the field,
${\cal P}_a = \sigma_a^2$.
So the root mean square fluctuations of the axion field and
initial misalignment angle are related to the Hubble constant by
\beq
\sigma_{a} = T_{\rm GH} = {\Hinf \over 2 \pi}
~~~\Rightarrow~~~ \sigma_{\theta} = {\Hinf \over 2 \pi (f_a / N)}
\label{sigmaHf}
\eq
where $T_{\rm GH}$ is the Gibbons-Hawking temperature \cite{Bunch}.
Numerically, the variance of the initial misalignment angle is
related to the Hubble constant during inflation and Peccei-Quinn scale by
\beq
\sigma_{\theta} \simeq 1.6 \times 10^{-4} \left( { \Hinf \over 10^{13}~{\rm GeV} }
  \right)
  \left( { 10^{16}~{\rm GeV}  \over f_a/N } \right)
\label{eq:variancebound}
\eq
This inflationary produced
variance provides a lower bound on the combination
$\theta_i^2+\sigma_\theta^2$ that appears in the expressions for the
axion relic density.

%(\ref{relicna}),

In order to establish a lower limit for the observational effects
of relic axion fluctuations in the CMBR
discussed in the next section,
it is useful to determine the minimum relic axion density as a function
of Hubble scale during inflation
for each late time cosmological scenario
described in the previous subsections.
Consider first the case in which the axion condensate forms
during a radiation dominated era without any dilution by a
late decaying particle.
For a Peccei-Quinn scale of
$f_a/N \lsim 6 \times 10^{17}$ GeV,
the relic density (\ref{relicdensity})
with $C=0.018$ and $\lqcd = 200$ MeV, and
for an average
initial misalignment angle which is small compared with the
variance induced by inflationary fluctuations
(\ref{eq:variancebound}), namely $\theta_i^2 \lsim \sigma_{\theta}^2$,
is bounded by
\beq
\Omega_a h^2 \gsim 5 \times 10^{-4} \left( { \Hinf \over 10^{13} ~{\rm GeV} }
\right)^2 \left( { 10^{16}~ {\rm GeV} \over f_a/N } \right)^{5/6}
~~~~~~{\rm for} ~~\theta_i^2 \lsim \sigma_{\theta}^2
\label{eqn:firstreldensity}
\eq
With the same late time cosmological scenario,
but with at least $f_a/N \gsim 6\times 10^{17}$ GeV,
%the axion
%oscillates with its zero temperature mass (see Eq.~\ref{relicdensitylargef}),
the relic density (\ref{relicdensitylargef})
with $f_c \xi(T_{\rm osc})^{-1} \simeq 1$ and
with small average initial misalignment angle,
$\theta_i^2 \lsim \sigma_{\theta}^2$, gives
a bound on the relic axion density of
\beq
\Omega_a h^2 \gsim 1 \times 10^{-4}
 \left( { \Hinf \over 10^{13} ~{\rm GeV} } \right)^2
 \left( { 10^{16}~ {\rm GeV} \over f_a/N } \right)^{1/2}
~~~~~~~{\rm for} ~~\theta_i^2 \lsim \sigma_{\theta}^2
\label{tndlargef}
\eq
The lower bounds (\ref{eqn:firstreldensity}) and (\ref{tndlargef})
for small and large Peccei-Quinn scales respectively
depend on {\it inverse} powers of $f_a/N$ because the variance
in initial misalignment angle
(\ref{sigmaHf}) is inversely proportional to the Peccei-Quinn scale.
And both scale like the square of the Hubble constant during inflation
since the relic density scales like the square variance of
axion field, and the variance scales like the Hubble constant.

Next consider the case of a late entropy release
in which the axion condensate is formed
during a matter dominated era with a later transition to
standard radiation domination by reheating from a massive particle.
In this case the relic density (\ref{rhbound})
for an average
initial misalignment angle which is small compared with the
variance induced by inflationary fluctuations
(\ref{eq:variancebound}), namely $\theta_i^2 \lsim \sigma_{\theta}^2$,
is bounded by
\beq
\Omega_a h^2 \gsim
  %8.7
  9 \times 10^{-7}
 \left( { T_{\rm RH} \over 6~{\rm MeV} } \right)^2
 \left( { \Hinf \over 10^{13}~ {\rm GeV} } \right)^2
~~~~~~~{\rm for} ~~\theta_i^2 \lsim \sigma_{\theta}^2
\label{eqn:lastreldensity}
\eq
Note that this bound on the residual relic axion density induced
by inflationary fluctuations is {\it independent} of the
Peccei-Quinn scale
because in this case the relic density (\ref{rhbound}) is proportional to
$(f_a/N)^2$ while the variance in initial misalignment angle
(\ref{sigmaHf}) squared is proportional to $(f_a/N)^{-2}$.

The bound (\ref{eqn:lastreldensity}) is only applicable
if the axion condensate is formed during the matter dominated
era before the dominating particle decays, which
requires $3H(T_{\rm RH})\lsim m_a(T_{\rm RH})$.
For a small enough axion mass the axion condensate is formed
in the radiation dominated era subsequent to the late particle
decay.
This will occur for $3H(T_{\rm RH}) \gsim m_a(T_{\rm RH})$
corresponding to a Peccei-Quinn scale
\beq
f_a/N\gsim 2 \times 10^{20}
   \left(\frac{6 {\rm \, MeV}}{T_{\rm RH}}\right)^2 ~\rm GeV
\label{eqn:changeover}
\eeq
%where $g_*\approx 10-17$ for $1\ {\rm MeV}\lsim T_{RH}\lsim \lqcd$.
In this extreme case the relic density (\ref{relicdensitylargef}) for
formation of the axion condensate in the zero temperature potential
in a radiation dominated era is applicable.
And the bound (\ref{tndlargef}) on the relic axion density from
inflationary induced axion fluctuations is obtained.
Such a large Peccei-Quinn scale would seem
unnatural and does not occur in any known string/M-theory
compactifications \cite{Banks:2003sx}.
However, it is useful to consider this regime in order
to assess the strength of the bounds described in section 4.

%Notice that for $f_a/N \sim 10^{16}$ GeV appropriate to a
%string/M-theory QCD axion and Hubble constant during inflation

% ------------------------------------------
% ------------------------------------------

\section{Axion Signatures in the CMBR}
\label{sec:Sigs}

A string/M-theory QCD axion with $f_a/N \sim 10^{16}$ GeV
is likely to be practically impossible to detect in the laboratory.
However, inflationary produced fluctuations of such an axion
do give rise to two distinctive signatures in CMBR temperature
fluctuations which are in principle measurable.
The first is an isocurvature component of temperature
fluctuations \cite{AxionIso}
which are discussed in section \ref{sec:isocurvature}.
The second is a non-Gaussian component of temperature
fluctuations \cite{lythstewart} discussed in section \ref{sec:nongauss}.

Both of these signatures depend on the relic axion density.
The minimum residual relic densities arising from inflationary axion
fluctuations presented in section \ref{infaxionsection}
are therefore a crucial ingredient in establishing lower bounds
for the observable isocurvature and non-Gaussian effects
in the CMBR temperature fluctuations.
These lower bounds, however, depend on the Hubble constant
during inflation.
So it is necessary to have some independent measurement which can
in principle establish the Hubble constant during inflation.
This can in fact be accomplished by a measurement of
primordial inflationary gravity wave contributions to CMBR temperature
fluctuations as described section \ref{gwaves} \cite{AbbottWise}.

%If the axion relic density is too small these two signatures in the
%CMBR will be unmeasurable.  As we saw in the previous section there is
%a lower bound on the axion relic density as a function of $H_{inf}$.
%Thus is will be very important to first measure a high Hubble constant
%during inflation in order that the effects of the axion be
%observable.  We review in Sec. \ref{gwaves} how the size of the
%Hubble constant during inflation can be measured from observation of
%gravity waves in the CMBR \cite{AbbottWise}.

%Other than the assumption that in the future we will observe these
%tensor modes in the CMBR we make no further assumptions.  No
%assumptions are made about the specific mechanism for possible small
%misalignment angle, origin of a possible diluting entropy release, or
%fraction of CDM which is comprised of relic axions.  In this way the
%extent to which precision cosmological measurements can truly probe a
%QCD string/M-theory axion may be addressed nearly independent of any
%cosmological assumptions.

% ------------------------------------------------

\subsection{Gravity Waves}
\label{gwaves}

All massless fields
undergo quantum de Sitter fluctuations during inflation,
as discussed in section \ref{infaxionsection}.
This includes the metric as well as an axion.
So a measurement of the magnitude of
metric fluctuations generated during
inflation would establish the magnitude of
the canonically normalized
axion fluctuations generated during inflation.
And for a given Peccei-Quinn scale
this would then establish a lower limit
for the magnitude of observable effects of a
string/M-theory QCD axion in CMBR temperature
fluctuations described in the following subsections.

The power spectral density of
inflationary dimensionless metric fluctuations may be written
\beq
\Power_{h}(k) \equiv {  k^3  \over  2 \pi^2}  2 \langle | h_{+ {\bf k}} |^2
  + | h_{ \times {\bf k}} |^2  \rangle
  \label{powerhdef}
\eq
where $h_+$ and $h_{\times}$ are the two polarization modes of the
metric perturbations.
%Up to an overall normalization this is identical to the power
%spectral density of axion fluctuations (\ref{apowerdensity}).
This power is related to the Hubble constant during inflation
by \cite{wmapinf},
\beq
\Power_h =2 \left( {2 \over M_p} { \Hinf \over 2 \pi} \right)^2
\label{hpower}
\eq
This is identical to the power
spectral density of axion fluctuations (\ref{poweraH})
up to a normalization related to canonical normalization
of metric fluctuations and overall normalization
chosen in (\ref{powerhdef}).

Metric perturbations generated during inflation appear
as gravity waves re-entering the horizon after inflation.
These gravity waves induce tensor B-modes in
CMBR temperature fluctuations.
In order to characterize the magnitude of the power
in gravity waves
it is useful to define the ratio of tensor to scalar
metric perturbations.
%modes in the CMBR temperature fluctuations.
Here we adopt the normalization used by the WMAP collaboration
\cite{wmapinf,llmds} in which the tensor to scalar ratio is defined by
the ratio of spectral power densities at some
reference scale $k_0$,
\beq
r = { \Power_h(k_0) \over \Power_{\cal R}(k_0) }
\label{rdefpower}
\eq
where
\beq
\Power_{\cal R}(k) \equiv  { k^3 \over 2 \pi^2}  \langle | {\cal R}_{\bf k}
  |^2 \rangle
\eq
is the power spectral density in scalar metric perturbations and
${\cal R}$ is the gauge invariant
dimensionless curvature perturbation
\cite{llbook}.
It is worth noting that the definition (\ref{rdefpower})
has the advantage that
the relation between $r$ and the Hubble constant during
inflation does not depend on cosmological parameters
such as the present dark matter or dark energy fractions
(in contrast to some other definitions of the tensor to scalar
ratio in the literature).
The scalar curvature perturbation spectral power density
measured from CMBR temperature fluctuations by WMAP
under the assumption that they arise predominantly from
scalar metric fluctuations is \cite{wmapinf}
\beq
\Power_{\cal R}(k_0) \simeq 2.1 \times 10^{-9}
~~~~~~{\rm at}~~ k_0=0.002 ~{\rm Mpc}^{-1}
\eq
With this, the variance of inflationary induced fluctuations
in the initial QCD misalignment angle (\ref{eq:variancebound})
may be related to the
scalar to tensor ratio (\ref{rdefpower}) using (\ref{hpower}) by
\beq
\sigma_{\theta} \simeq 1.7 \times 10^{-3} ~ \sqrt{r} ~
\left( { 10^{16}~{\rm GeV} \over f_a/N } \right)
\label{sigmarrel}
\eq
For a given Peccei-Quinn scale this gives a relation between
an observable quantity and the variance of the
initial misalignment angle.

The current
WMAP bound on the scalar to tensor
ratio from the non-observation of B-mode tensor components in
CMBR temperature fluctuations
is $r \lsim 1.3$ \cite{wmapinf}.
With the definition (\ref{rdefpower}) and (\ref{hpower})
this corresponds to an upper limit
on the Hubble constant during inflation of
$H_{\rm inf} \lsim 2.8 \times 10^{14}$ GeV.
Future experiments will be capable of probing significantly smaller
values of the scalar to tensor ratio $r$ and concomitantly
smaller values of the Hubble constant during inflation.
The PLANCK polarimetry experiment could ultimately be sensitive
at the one sigma level to
%$r \gsim 7 \times 10^{-3}$ corresponding
%to $\Hinf \gsim 2 \times 10^{13}$ GeV
at best roughly $r \gsim 5 \times 10^{-4}$ corresponding
to $H_{\rm inf} \gsim 5 \times 10^{12}$ GeV \cite{InflationEdges,rpref}.
This represents a very optimistic sensitivity for a
positive observation.
So throughout we will
consider $H_{\rm inf} \gsim 10^{13}$ GeV as a reasonable benchmark
for a positive observation of primordial gravity waves
by the PLANCK polarimetry experiment.
If achieved this would represent a significant improvement
over the current WMAP bound.
If PLANCK does indeed observe tensor B-modes at any level
between the current WMAP bound and its ultimate sensitivity,
this would imply a lower bound on the axion relic density
for a given Peccei-Quinn scale in
the various cosmological scenarios discussed in section 2.4.
%(\ref{eqn:firstreldensity})-(\ref{eqn:lastreldensity}).
The ultimate sensitivity to primordial gravity waves
achievable in future CMBR experiments is limited
not only by issues of instrumentation, but also from the fact that
tensor B-mode fluctuations can be generated at second order
in primordial scalar fluctuations from gravitational lensing effects.
This provides an irreducible background for {\it any} future
probe of primordial gravity waves in CMBR experiments.
In terms
of the scalar to tensor ratio this has been
estimated to be roughly $r \gsim 7 \times 10^{-5}$
corresponding to $H_{\rm inf} \gsim 2 \times 10^{12}$ GeV
\cite{InflationEdges,knoxsong}.

% -----------------------------------------------------

\subsection{Isocurvature Perturbations}
\label{sec:isocurvature}

Quantum fluctuations generated during inflationary
can lead to observable signatures in CMBR temperature
fluctuations in a number of ways.
Fluctuations of the inflaton field give rise to fluctuations
in the scalar curvature and lead to adiabatic fluctuations
in the CMBR.
Fluctuations of the metric lead to tensor B-model fluctuations
in the CMBR, as discussed above.
Inflationary induced fluctuations of an axion field
turn out to lead isocurvature fluctuations in the CMBR \cite{AxionIso}
as reviewed below.
This distinctive signature provides a probe for the existence
of an axion field during inflation.

Isocurvature perturbations correspond to fluctuations in the local
equation of state of some species, $\delta (n_i/s)\neq 0$,
with no fluctuation in the total energy density, $\delta \rho=0$.
These are in some sense orthogonal to adiabatic
perturbations which correspond to fluctuations in the total
energy density, $\delta \rho \neq 0$,  with no fluctuation
in the local equation of state, $\delta (n_i/s) =0$.
During inflation the potential for a QCD axion is essentially flat
as long as the Peccei-Quinn symmetry remains unbroken during this
epoch, as discussed above.
Fluctuations in the axion field therefore do not affect the
total density and are isocurvature.

In order to discuss both
adiabatic and isocurvature fluctuations it is useful to define
the fractional fluctuation in the number density of
the $i$-th species divided by the entropy density
\beq
S_i \equiv { \delta (n_i/s) \over n_i/s } =
{\delta n_i \over n_i} - 3 { \delta T \over T}
\label{Ssdef}
\eq
where the second equality follows from $s \propto T^3$.
Throughout, all fields
other than the axion are assumed to undergo adiabatic fluctuations,
$S_i=0$ for $i \neq a$.
Relaxing this assumption would only allow for a smaller
axion contribution for a fixed total isocurvature
contribution to CMBR temperature fluctuations, and so could
only result in stronger bounds than those given in section 4.
As described above, isocurvature fluctuations of the axion
on length scales larger than the horizon do not modify the total
density on these scales, $\delta \rho_{\rm iso} \simeq 0$.  After the
axion condensate forms, the total energy density is a sum over
non-relativistic species including the axion plus radiation
\beq
\rho = \sum_i m_i n_i + m_a n_a +
\rho_r
\eq
where $\rho_r \propto T^4$ is the radiation energy density,
and throughout $\sum_i$ is over all non-relativistic species
but the axion.
The number density and temperature
fluctuations for an isocurvature fluctuation
on scales large compared with
the horizon in this epoch are then related by
\begin{equation}
\delta \rho_{\rm iso}=
 \sum_{i} m_i \delta n_i + m_a \delta n_a + 4 \rho_{r}
 \frac{\delta T}{T} \simeq 0
\label{drhoiso}
\end{equation}
Since all species other than the axion are assumed to fluctuate
adiabatically, $S_i=0$, the definition (\ref{Ssdef}) relates the
number density fluctuation in each species to the initial isocurvature
temperature fluctuation by $\delta n_i = 3 n_i ( \delta T / T)$.
With this, the isocurvature constraint (\ref{drhoiso})
then relates the isocurvature
temperature fluctuation to the axion number density fluctuations after
the axion condensate is formed
\begin{equation}
\frac{\delta T}{T}
  \simeq - {\rho_a \over 3\sum_{i} \rho_i + 4\rho_{r} }
\frac{\delta n_a}{n_a}
\label{dtdn}
\end{equation}
where $\delta \rho_a = m_a \delta n_a$.
The relation (\ref{Ssdef})
may then be used to write the axion induced temperature fluctuation
on super-horizon scales in
terms of the fractional axion fluctuation $S_a$
\begin{equation}
\frac{\delta T}{T}
 \simeq  - {\rho_a \over 3( \sum_{i} \rho_i + \rho_a) + 4\rho_{r} }
  ~S_a
\label{dTsrelation}
\end{equation}
This relation is more useful than (\ref{dtdn}) for arbitrary
times after the axion condensate forms since
$S_a$ rather than $\delta n_a/ n_a$ is approximately constant
for fluctuations on super-horizon size scales \cite{KolbTurner}.

The fractional axion fluctuation $S_a$ can be related
directly to the inflationary induced fluctuations in the axion field.
When the axion condensate is formed at a temperature of $T_{\rm osc}
\sim \Lambda_{\rm QCD}$, in any of the cosmological scenarios
discussed in section 2, it comprises a small fraction of the total
energy density, $\rho_a \ll \sum_i \rho_i + \rho_r$.  From
(\ref{dtdn}) it follows that just after the axion condensate is
formed, the axion induced fractional temperature fluctuations are much
smaller than the fractional axion number density fluctuations,
$(\delta T / T)_{\rm init} \ll (\delta n_a / n_a)_{\rm init}$.
From this it follows that the
fractional fluctuation $S_a$ defined in (\ref{Ssdef}) just after the axion
condensate is formed may then be related to fluctuations in the
initial misalignment angle
\beq
S_a \simeq { \delta n_a \over n_a}_{\rm init}
  \simeq {\delta (\theta^2) \over
   \langle \theta^2 \rangle }=
   \frac{(\av{\theta} + \delta\theta)^2-\av{\theta^2}}{\av{\theta^2}}=
{2 \theta_i \delta\theta+( \delta \theta)^2
  -\sigma_{\theta}^2 \over  \theta_i^2+\sigma_{\theta}^2 }
\label{Sdtrelation}
\eq
where $\delta \theta$ and $\sigma_{\theta}$ are understood to be
due to the initial inflationary induced fluctuations in the misalignment
angle, $\theta_i$ is the average misalignment angle over the currently
observable universe, and $n_a \propto \theta^2$ ignoring anharmonic effects
(this is a very good assumption for $f_a/N \sim 10^{16}$ GeV
since $\theta \ll \pi$ is required in this case).
And since $S_a$ is approximately constant for
super-horizon size fluctuations \cite{KolbTurner}, the
relation (\ref{Sdtrelation}) between $S_a$
and the initial misalignment angle fluctuations remains good for
super-horizon modes for arbitrary times after the axion condensate forms.
For later reference, with the relation (\ref{Sdtrelation}),
the mean square of the fractional axion fluctuation
on super-horizon scales %, $\langle S_a^2 \rangle$,
is related to the mean square fluctuations of, and average value of,
the initial misalignment angle %, $\sigma_{\theta}^2$ and $\theta_i$,
by
\beq
%\av{ S_a^2 }
 \langle S_a^2 \rangle
   = 2 \sigma_{\theta}^2
  { 2 \theta_i^2 + \sigma_{\theta}^2 \over
    ( \theta_i^2 + \sigma_{\theta}^2 ) ^2 }
\label{Sthetarel}
\eq
where $\delta \theta = \theta- \langle \theta \rangle$ is assumed to be
Gaussian distributed, $\theta_i = \langle \theta \rangle$, and
$\sigma_{\theta}^2 = \langle ( \theta - \langle \theta \rangle )^2 \rangle$.

As discussed below, the most relevant axion induced temperature
fluctuations are those on the largest angular scales.
These temperature fluctuations enter the horizon
well into the matter dominated era for which
$\sum_i \rho_i +\rho_a \gg \rho_r$.
So from (\ref{dTsrelation}) the axion induced initial temperature
fluctuations on large angular scales at horizon crossing are
related to the axion isocurvature fluctuation $S_a$ by
\beq
{ \delta T \over T}_{\rm horizon} \simeq - {1 \over 3}
{\Omega_a \over \Omega_m}
  S_a
\label{dTinit}
\eq
where $\Omega_a = \rho_a / \rho_c$ is the fractional axion density
and $\Omega_m = (\sum_i \rho_i + \rho_a)/ \rho_c$
is the fractional density of
all non-relativistic matter including baryons, axions, and
any other species which contribute to cold dark matter,
$\Omega_m = \Omega_{\rm CDM}+\Omega_b$.
%and $S_a$ is related to
%fluctuations in the initial misalignment angle by (\ref{Sdtrelation}).
From WMAP the total matter density in a $\Lambda$CDM cosmology
is measured to be \cite{WMAP}
\beq
\Omega_m h^2= 0.135^{+0.008}_{-0.009}
\label{wmapomegam}
\eq
An experimental observation of temperature fluctuations corresponds to
the initial fluctuation at horizon crossing (\ref{dTinit})
plus an additional Sachs-Wolfe contribution
\cite{Sachs} from the integrated redshift between the surface of last scattering
and detector.  The latter contribution turns out to be five times
smaller \cite{Bond} than that from the initial temperature fluctuation
(\ref{dTinit}),
\beq
{ \delta T \over T}_{\rm SW} \simeq - {1 \over 15}
{\Omega_a \over \Omega_m}
% {\delta n_a \over n_a}
 S_a
\eq
The total observable isocurvature temperature fluctuation is then related
to the axion isocurvature fluctuation by
\beq
{ \delta T \over T}_{\rm iso} \simeq - {6 \over 15}
{\Omega_a \over \Omega_m}
%{\delta n_a \over n_a}
 S_a
\label{dTiso}
\eq

%%%%%%%%%%%%%%%%%%%%%%%%%%%%%%%%%55
\sfig{cmbpower}{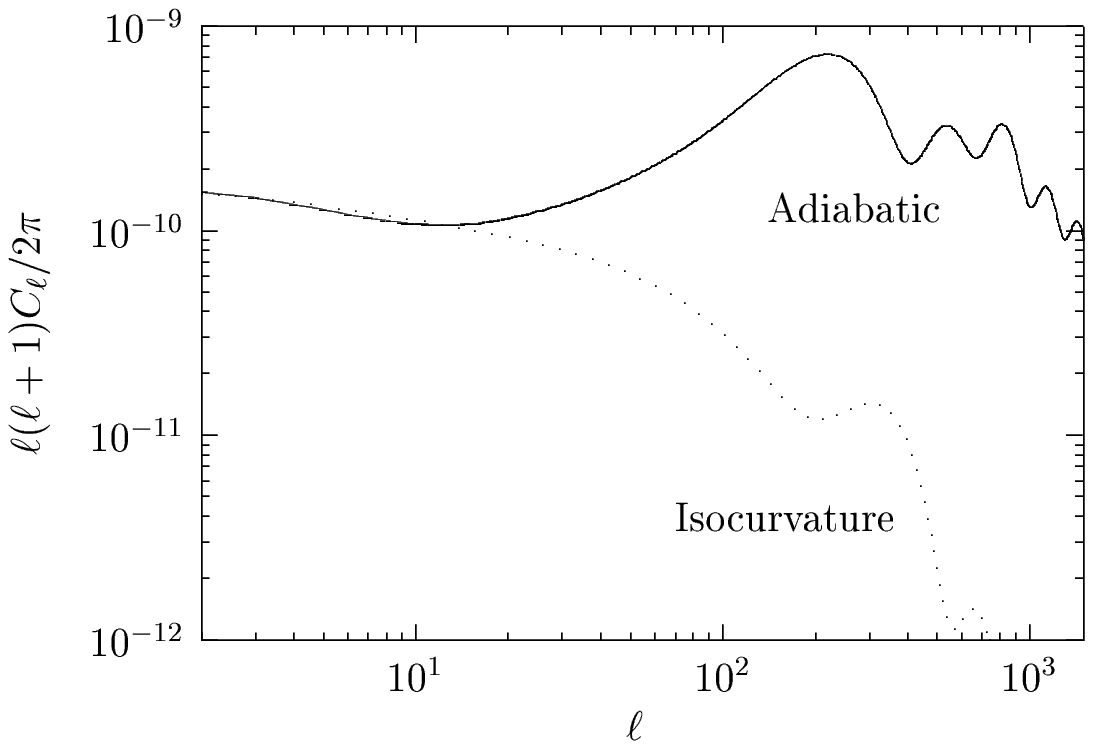}{169}{342}{482}{555}{4.75in}{0in}{
Temperature fluctuation power spectrum with pure adiabatic
(solid line) or pure isocurvature (dashed line) initial fluctuations,
$\Delta T(\theta,\phi) / T = \sum_{\ell m} a_{\ell m} Y_{\ell m}(\theta,\phi)$,
$C_{\ell} = \langle | a_{\ell m} |^2 \rangle$.
$\Lambda$CDM cosmology with $\Omega_{\Lambda} = 0.73$,
$\Omega_{\rm CDM} = 0.22$, $\Omega_{b} = 0.05$,
$H=72 ~{\rm km}~s^{-1}~{\rm Mpc}^{-1}$,
and spectral index $n=1$.}
%%%%%%%%%%%%%%%%%%%%%%%%%%%%%%%%%%%%%%

The expression (\ref{dTiso}) relating axion induced temperature
fluctuations with the fractional axion fluctuation applies
strictly only on the largest angular scales.
The spectrum of temperature fluctuations is modified
substantially on smaller scales corresponding to multipoles
well beyond the Sachs-Wolfe plateau.
%by sub-horizon processes.
This is illustrated in Fig.~\ref{cmbpower} which gives the CMBR temperature
fluctuation power spectrum for pure adiabatic or pure
isocurvature initial fluctuations in a $\Lambda$CDM cosmology
generated by CMBFAST version 4.3 \cite{cmbfast}.
The power in isocurvature fluctuations falls only by a factor of
two for roughly $\ell \sim 50$ but continues to fall
rapidly for larger $\ell$.
This rapid decrease in isocurvature contributions to
the power in CMBR temperature fluctuations compared with
adiabatic contributions arises for two reasons.
First, the growth of sub-horizon size isocurvature and adiabatic temperature
fluctuations are completely out of phase with one another.
So while the magnitude of adiabatic fluctuations are
growing towards the first peak, the magnitude of isocurvature
fluctuations are falling towards a first minimum.
Secondly, fluctuations
%beyond roughly the first isocurvature minimum
well beyond the Sachs-Wolfe plateau
enter the horizon during radiation domination.
Unlike adiabatic fluctuations, the temperature component of
isocurvature
fluctuations during this epoch are suppressed by the radiation
density, as can be seen in the denominator of (\ref{dTsrelation}).
This follows since well back into
the radiation dominated era the isocurvature
condition (\ref{drhoiso}) %since $\rho_{\rm iso} \simeq 0$
can be satisfied for fixed magnitude of axion density fluctuation
with vanishing small temperature fluctuation.
This fall off of isocurvature temperature fluctuations
with multipoles beyond the Sachs-Wolfe plateau
is important in considering how well such fluctuations
can be probed either by a direct measure of the
power spectrum or search for non-Gaussian components discussed
in the next subsection.

In order to characterize the observable effects of
isocurvature fluctuations it is useful to define the ratio
of average power in the isocurvature
component to the average total power in
CMBR temperature fluctuations
%\beq
%\alpha \equiv \frac{\av{\left(\frac{\delta
%        T}{T}\right)_{{\rm iso}}^2}}{\av{\left(\frac{\delta
%        T}{T}\right)_{\rm tot} ^2}}
%\label{alphadef}
%\eq
\beq
\alpha \equiv \frac{\av{\left( \delta T/ T \right)_{{\rm iso}}^2}}
    {\av{\left( \delta T/ T \right)_{\rm tot} ^2}}
\label{alphadef}
\eq
The total root mean square CMBR temperature fluctuation
was measured by COBE to be \cite{COBE}
\beq
%\av{ \left( { \delta T \over T} \right)_{\rm tot} ^2 }^{1/2}
 \av{ \left( { \delta T / T} \right)_{\rm tot} ^2 }^{1/2}
% \langle \left( { \delta T / T} \right)_{\rm tot} ^2 \rangle^{1/2}
 \simeq 1.1 \times 10^{-5}
\label{cobedt}
\eq
Because of the rapid fall off of isocurvature power, the ratio
(\ref{alphadef}) receives contributions mainly from low multipoles
on the Sachs-Wolfe plateau.
And since, as discussed above, axion induced isocurvature temperature
fluctuations are fairly well approximated by (\ref{dTiso}) for these low
multipoles, the isocurvature power ratio (\ref{alphadef})
may be approximately related to the mean square fluctuations
and average value of the initial misalignment angle using (\ref{Sthetarel})
by
%\begin{equation}
%\alpha
%       \simeq  \left( { 6 \over 15} \right)^{2}
%       \frac{\left(\frac{\Omega_a}{\Omega_m}
%               %{\delta n_{a} \over n_a}
%           \right)^2
%      \langle S_a ^2 \rangle
%    }{\av{ \left( \frac{\delta T}{T} \right)_{\rm tot}^{2} } }
%        \simeq
%  \left( { 6 \over 15}
%      \right)^{2}\frac{\left(\frac{\Omega_a}{\Omega_m}\right)^2
%}{ \av{ \left(\frac{\delta T}{T} \right)_{\rm tot}^{2}  }  }
%      2\sigma_\theta^2\frac{2\theta_i^2+\sigma_\theta^2}
%         {\left(\theta_i^2+\sigma_\theta^2\right)^2}
%       \label{alphasa}
%\end{equation}
\begin{equation}
\alpha
       \simeq  \left( { 6 \over 15} \right)^{2}
       \frac{\left( \Omega_a / \Omega_m
               %{\delta n_{a} \over n_a}
           \right)^2
      \langle S_a ^2 \rangle
    }{\av{ \left( \delta T / T \right)_{\rm tot}^{2} } }
         \simeq
  \left( { 6 \over 15}
      \right)^{2}\frac{\left(\Omega_a / \Omega_m\right)^2
}{ \av{ \left(\delta T/ T \right)_{\rm tot}^{2}  }  }
      ~2\sigma_\theta^2\frac{2\theta_i^2+\sigma_\theta^2}
          {\left(\theta_i^2+\sigma_\theta^2\right)^2}
        \label{alphasa}
\end{equation}
A full calculation of the relation between $\alpha$ and
%$\langle S_a^2 \rangle$
the variance and average value of the initial misalignment angle
utilizing the full isocurvature power spectrum illustrated
in Fig. 1 would not differ significantly from the approximation
(\ref{alphasa}).

%For the bounds presented in section 4,

It is instructive to
consider the numerical relation (\ref{alphasa}) between the axion fluctuation
induced isocurvature power fraction and the variance and average
initial misalignment angle in each of the
cosmological scenarios for formation of the axion condensate
discussed in section 2.
For the case of smaller values of the
Peccei-Quinn scale in which the axion condensate forms during a radiation
dominated era in the high temperature axion potential and
without any dilution after its formation, the
relic axion density is given by (\ref{relicdensity})
in section 2.1.
This relic density with $C=0.018$ and $\lqcd=200$ MeV and ignoring
anharmonic effects, along with the WMAP total
matter density measurement (\ref{wmapomegam}) and COBE
total root mean square CMBR temperature fluctuation measurement
(\ref{cobedt}) gives a fractional isocurvature power (\ref{alphasa})
of
\beq
\alpha \sim 6 \times 10^{19}
\left( { f_a/N \over 10^{16}~{\rm GeV} } \right)^{7/3}
~ \sigma_{\theta}^2 \left( 2 \theta_i^2 + \sigma_{\theta}^2 \right)
\label{alphalowf}
\eq
For the case of larger Peccei-Quinn scales
in which axion condensate forms during a radiation
dominated era but in effectively the zero temperature axion potential and
again without any dilution after its formation, the
relic axion density is given by (\ref{relicdensitylargef})
in section 2.1.
This relic density with $f_c \xi(T_{\rm osc})^{-1} \simeq 1$ and ignoring
anharmonic effects, along with
(\ref{wmapomegam}) and (\ref{cobedt})
gives a fractional isocurvature power (\ref{alphasa}) of
\beq
\alpha \simeq 3.6 \times 10^{18}
\left( { f_a/N \over 10^{16}~{\rm GeV} } \right)^{3}
~ \sigma_{\theta}^2 \left( 2 \theta_i^2 + \sigma_{\theta}^2 \right)
\label{alphahighf}
\eq
Finally, for the case in which the axion condensate forms during
a matter dominated era with subsequent
late decay of the dominating particle with a reheat temperature
$T_{\rm RH} \lsim \lqcd$, the relic density is given by (\ref{rhbound}) in
section 2.3.1.
This relic density with $f_c \xi(T_{\rm osc})^{-1} \simeq 1$ and ignoring
anharmonic effects, along with
(\ref{wmapomegam}) and (\ref{cobedt})
gives a fractional isocurvature power (\ref{alphasa}) of
\beq
\alpha \simeq 1.8 \times 10^{14}
\left( { T_{\rm RH} \over 6~{\rm MeV} } \right)
\left( { f_a/N \over 10^{16}~{\rm GeV} } \right)^{4}
~ \sigma_{\theta}^2 \left( 2 \theta_i^2 + \sigma_{\theta}^2 \right)
\label{alphatrf}
\eq
Note that the fractional isocurvature power is strictly bounded
by $\alpha \leq 1$.
So for a string/M-theory axion with $f_a/N \sim 10^{16}$ GeV
the large pre-factors in (\ref{alphalowf})--(\ref{alphatrf})
imply that
even just the magnitude of the observed temperature fluctuations
(\ref{cobedt}) places a stringent limit on the combination
$\sigma_{\theta}^2 ( 2 \theta_i^2 + \sigma_{\theta}^2 )$
of variance and average value of the initial misalignment angle
in any cosmological scenario for the formation of the axion
condensate in which the axion exists during inflation.

The isocurvature power
ratio (\ref{alphadef}) may be measured or bounded by fitting the
observed CMBR temperature fluctuation
power spectrum with a linear combination of adiabatic
and isocurvature components with some assumptions about the
underlying cosmological model.
%including the tilt of the primordial fluctuations.
Using WMAP data and an analysis
in which the tilt of both the adiabatic and isocurvature
components are allowed to vary over fairly large ranges,
the constraint $\alpha \lsim 0.4$ has been obtained \cite{GarciaBellido}.
This analysis is very conservative and weakens
the bound due to some degeneracies in parameters.
The magnitude of inflationary induced axion isocurvature fluctuations
depends on the Hubble constant during inflation.
And in most theories of inflation the Hubble constant during
inflation is by definition a very slowly changing quantity.
This results in a nearly scale invariant spectrum of axion induced
isocurvature temperature fluctuations, $n_{\rm iso} \simeq 1$.
If this prior were imposed, the resulting bound on
$\alpha$ using WMAP data would likely be at the many percent
level.
Since most of the power in isocurvature
temperature fluctuations is restricted to low multipoles, future
measurements will not improve knowledge of the power
spectrum there since the current measurements are already cosmic variance
limited even beyond the first peak.
However, future measurements at higher multipoles along with
additional information from polarization, such as from PLANCK,
will resolve degeneracies in the power spectrum such as the
adiabatic tilt.
And this may well allow the fractional isocurvature power $\alpha$
to be bounded at the percent level.
It is worth noting that gravity waves also contribute to
CMBR temperature mainly at low multipoles on the Sachs-Wolfe plateau.
If gravity waves are indeed detected in
a future CMBR experiment, then
even less of the budget of observed power at low multipoles
can be due to a isocurvature component which would
further strengthen the bound on $\alpha$.

% -------------------------------------------------------

\subsection{Non-Gaussianity}
\label{sec:nongauss}

Primordial inflationary fluctuations in a QCD axion field give rise
to CMBR temperature fluctuations.
These fluctuations have the distinctive feature of
being isocurvature \cite{AxionIso} with an angular power spectrum
which differs from that of adiabatic fluctuations, as described in the
previous subsection.
Axion induced CMBR temperature fluctuations turn out also to
have a non-Gaussian component \cite{lythstewart}
as reviewed below.
This additional distinctive feature provides another probe
for the existence of a QCD axion field during inflation.

Quantum fluctuations of a free massless field generated during inflation
are Gaussian distributed.
Slow roll inflation generally requires an inflaton with exceedingly small
self coupling and so its fluctuations act essentially like a free field.
Inflaton fluctuations in most models are therefore Gaussian to
a very high degree.
The adiabatic temperature fluctuations which
ultimately result from inflaton induced fluctuations of the
scalar curvature are then also Gaussian distributed.
As discussed above, as long as the Peccei-Quinn symmetry
remains unbroken, a QCD axion is also essentially a free massless
field during inflation and so the initial misalignment angle
undergoes Gaussian distributed fluctuations.
However, the axion induced CMBR temperature fluctuations
are proportional to fluctuations in the axion number density
when the axion condensate is formed, as discussed in section 3.2.
But the number density is proportional to the square of the
misalignment angle ignoring anharmonic effects,
$n_a \propto \theta^2$, while it is
the fluctuations in the misalignment angle, $\delta \theta$,
which are Gaussian distributed.
This non-linearity introduces a non-Gaussian component in the
axion induced temperature fluctuations.

On large angular scales the axion induced temperature fluctuations
(\ref{dTiso}) may be related to fluctuations in the initial
misalignment angle through the relation (\ref{Sdtrelation}) as
\beq
{ \delta T \over T}_{\rm iso} \simeq - {6 \over 15}
{\Omega_a \over \Omega_m}
{2 \theta_i \delta\theta+( \delta \theta)^2
  -\sigma_{\theta}^2 \over  \theta_i^2+\sigma_{\theta}^2 }
  \label{dtisonon}
\eq
The first term in the numerator proportional to $\delta \theta$
is Gaussian distributed with zero mean, while the remaining terms
together, $(\delta\theta)^2-\sigma_\theta^2$, have a $\chi^2$
distribution with zero mean.
So primordial Gaussian distributed fluctuations in the
initial misalignment angle
lead in general to a combination of Gaussian and $\chi^2$
distributed isocurvature temperature fluctuations.
In the limit of large average misalignment angle compared
with the variance,
$\theta_i \gg\sigma_{\theta}$, the axion induced temperature fluctuations
(\ref{dtisonon}) are mostly Gaussian, whereas in the opposite limit,
$\theta_i \ll \sigma_{\theta}$, they are mostly $\chi^{2}$.
%In general, the
%contribution from the axion will be a linear combination of these two.
The relative importance of the Gaussian and $\chi^2$ fluctuations is
determined by the ratio $\theta_i / \sigma_{\theta}$.
The temperature fluctuation (\ref{dtisonon}) in these two limits is
\beq
{ \delta T \over T}_{\rm iso} \simeq - {6 \over 15}
{\Omega_a \over \Omega_m}
\left\{ \begin{array}{lll}
  {2 \delta \theta / \theta_i } & &
      \theta_i \gg \sigma_{\theta} \\
   (\delta \theta)^2/\sigma_{\theta}^2-1  & & \theta_i \ll \sigma_{\theta}
\end{array}
\right.
\eq
%In either limit  are suppressed by the relic axion
%density.
For $\theta_i \gg\sigma_{\theta}$ the magnitude of the axion
induced temperature fluctuations are
suppressed by the relic axion density and
additionally suppressed on average by $\sigma_{\theta} / \theta_i$.
However, for $\theta_i \ll \sigma_{\theta}$ they are only suppressed
by the relic axion density since the fractional axion fluctuations
are order one in this case.
The latter limit turns out to be most important in obtaining
the most conservative bounds discussed in section 4.

%It is instructive to consider the root mean square axion
%fluctuations (\ref{Sdtrelation}) in the limits in which either the Gaussian
%or $\chi^2$ distributions are dominant
%\beq
%\langle S_a^2 \rangle^{1/2} \simeq
%%\av{ \left( { \delta n_a \over n_a} \right)^2}^{1/2}
%\av{ \left( { \delta (\theta^2) \over \langle \theta^2 \rangle }
%    \right)^2}^{1/2}
%% \rightarrow
% \simeq
%\left\{ \begin{array}{ccl}
%  {2 \sigma_{\theta} / \theta_i } & &
%      \theta_i \gg \sigma_{\theta} \\
%   \sqrt{2}  & & \theta_i \ll \sigma_{\theta}
%\end{array}
%\right.
%\eq
%For $\theta_i \gg \sigma_{\theta} $ the fractional axion fluctuations are
%approximately Gaussian with magnitude determined by
%$\sigma_{\theta} / \theta_i$, while for
%$\theta_i \ll \sigma_{\theta} $ they are approximately $\chi^2$
%with order one magnitude.
%Eqns.~(\ref{eqn:firstreldensity})--(\ref{eqn:lastreldensity})
%indicate that the relic abundance of axions is substantially
%diminished in the regime $\theta_i \ll \sigma_{\theta}$.  This regime
%gives the most conservative constraints.

Gaussian fluctuations with zero mean are completely defined
in terms of the two point function.
All odd point functions vanish, and all even point functions are
products of the two point function.
So in order to test for non-Gaussianity it is necessary
to measure three or higher point functions.
In the present context the simplest possibility to characterize a
non-Gaussian component in CMBR temperature fluctuations
is to measure the average three point
function of the temperature fluctuation.
This is conveniently normalized as a dimensionless skewness
%\be
%S_{3,{\rm iso}} \equiv
%  \frac{\av{\left(\frac{\delta T}{T}\right)^3}}
%    {\av{\left(\frac{\delta
%      T}{T}\right)^2}^{3/2}},
%\label{eqn:skewness}
%\ee
\be
S_{3} \equiv
  \frac{\av{\left( \delta T / T \right)^3}}
    {\av{\left( \delta T / T \right)^2}^{3/2}}
\label{eqn:skewness}
\ee
For a pure Gaussian distribution the skewness vanishes while
for a pure $\chi^2$
distribution with zero mean $S_3 = - \sqrt{8}$.
Of course the full %the temperature fluctuation
three point function bi-spectrum contains additional information
which would be invaluable in discerning the origin of non-Gaussianity
in CMBR temperature fluctuations if ever observed.
However, in the absence of such positive measurements the
dimensionless skewness (\ref{eqn:skewness}) provides a good
dimensionless
normalized measure to characterize the magnitude of non-Gaussian components.

In order to evaluate the dimensionless skewness (\ref{eqn:skewness})
we consider the special case in which CMBR temperature
fluctuations are an uncorrelated sum of adiabatic plus
isocurvature contributions
\beq
{\delta T \over T} = {\delta T \over T}_{\rm ad} +
{\delta T \over T}_{\rm iso}
\label{dtdef}
\eq
This is the case if inflaton fluctuations are responsible
for the dominant adiabatic temperature fluctuations
observed in the CMBR, and the only other massless field during
inflation is a QCD axion.
Inclusion of gravity wave tensor B-mode contributions from inflationary
metric fluctuations would not modify the results below.
And inclusion of other sources of non-Gaussianity would only
allow for a smaller axion contribution for a fixed total
skewness of CMBR temperature fluctuations, and so could
only result in stronger bounds than those given in section 4.
Since the inflationary axion fluctuation induced isocurvature
temperature fluctuations have both Gaussian and $\chi^2$
components, the total temperature fluctuation (\ref{dtdef})
may be written in terms of a sum of distributions
\be
\frac{\delta T}{T}=\dt{I}+\dt{a}=\phi_I+\phi_a+\psi_a
\label{eqn:stats}
\ee
where $\phi_I$ parameterizes the Gaussian distributed
inflaton induced temperature fluctuations,
$\phi_a$ the Gaussian distributed component of the axion induced
temperature fluctuations, and
$\psi_a$ the $\chi^2$ distributed non-Gaussian component of the axion
induced temperature fluctuations.
Each distribution by definition has zero mean,
$\langle \phi_I \rangle =\langle \phi_a \rangle =\langle \psi_a \rangle=0$.
Note that $\phi_a$ and
$\psi_a$ are not independent since they both arise from axion
fluctuations whereas $\phi_I$ is independent of both $\phi_a$ and
$\psi_a$ since it arises from uncorrelated fluctuations of an
independent field.
So cross correlations between $\phi_I$ and $\phi_a$ or
$\psi_a$ vanish.
With this,
the two and three point temperature fluctuation correlation
functions from (\ref{eqn:stats}) are then
\be\label{eqn:twopoint}
\av{\frac{\delta T}{T}\frac{\delta T}{T}}=\av{\dt{I}\dt{I}}+\av{\dt{a}\dt{a}}=
\av{\phi_I\phi_I}+ \av{\phi_a\phi_a}
+\av{\psi_a\psi_a}
\ee
\be\label{eqn:threepoint}
\av{\frac{\delta T}{T}\frac{\delta T}{T}\frac{\delta T}{T}}=
\av{\dt{a}\dt{a}\dt{a}}= 3\av{\phi_a\phi_a\psi_a} +
3\av{\phi_a\psi_a\psi_a}+ \av{\psi_a\psi_a\psi_a}
\label{dtdtdt}
\ee
The three point function of course only depends
on the axion fluctuation induced isocurvature components
since by assumption the adiabatic inflaton fluctuations
are purely Gaussian distributed.

Now the dimensionless skewness defined in (\ref{eqn:skewness}) is an average
over all angular scales.
Axion isocurvature
contributions %to the skewness (\ref{eqn:skewness})
through the three-point function (\ref{dtdtdt}) however come mainly
from low multipoles on the Sachs-Wolfe plateau, just as for
the isocurvature power discussed in section 3.2.
On these large angular scales the relation between
the axion induced isocurvature temperature fluctuation and
fluctuations of %, variance in, and average value of
the initial misalignment angle is fairly well approximated by
(\ref{dtisonon}).
The relation between the axion induced
Gaussian and $\chi^2$ distributed temperature
fluctuation distributions $\phi_a$ and $\psi_a$ in (\ref{eqn:stats})
and the fluctuations in the initial misalignment angle
may then be approximated by
\begin{eqnarray}
\phi_a & \simeq & - {6 \over 15} \frac{\Omega_a}{\Omega_m}
 \frac{2\theta_i \delta\theta }{\theta_i^2+\sigma_{\theta}^2} \hspace{.5in}
\mbox{Gaussian}
\\
\psi_a & \simeq & - {6 \over 15}
  \frac{\Omega_a}{\Omega_m}
 { (\delta\theta)^2- \sigma_{\theta}^2 \over
  \theta_i^2+\sigma_{\theta}^2 } \hspace{.5in}
\mbox{$\chi^{2}$}
\end{eqnarray}
With this, the three point function (\ref{dtdtdt}) and dimensionless
skewness (\ref{eqn:skewness}) for the axion induced isocurvature
component of temperature fluctuations may be approximately related to the
mean square fluctuations and average value of the initial
misalignment angle by
\beq
  S_{3,{\rm iso}} \simeq -8 \left( {6 \over 15} \right)^3
 { (\Omega_a / \Omega_m)^3 \over
    \av{ \left( \delta T / T \right)_{\rm tot}^{2}  }^{3/2} }
          ~ \sigma_{\theta}^4
          { 3 \theta_i^2 + \sigma_{\theta}^2 \over
           (\theta_i^2 + \sigma_{\theta}^2)^3 }
\label{sapp}
\eq
A full calculation of the relation between $S_{3,{\rm iso}}$ and the
variance and average value of the initial misalignment angle
utilizing the full isocurvature power spectrum would
not differ significantly from the approximation (\ref{sapp}).
Note that this skewness is necessarily negative as a result
of the $\chi^2$ nature of the non-Gaussian component of axion
fluctuations.
This would be the first
test that a positive measurement of non-Gaussianity in CMBR
temperature fluctuations is in fact due to a QCD axion.

It is instructive to
consider the numerical relation between the axion fluctuation
induced isocurvature skewness (\ref{sapp}) and the variance and average
value of the initial misalignment angle in each of the
cosmological scenarios for formation of the axion condensate
discussed in section 2.
For the case of smaller values of the
Peccei-Quinn scale in which the axion condensate forms during a radiation
dominated era in the high temperature axion potential and
without any dilution after its formation, the
relic axion density is given by (\ref{relicdensity})
in section 2.1.
This relic density with $C=0.018$ and $\lqcd=200$ MeV and ignoring
anharmonic effects, along with the WMAP total
matter density measurement (\ref{wmapomegam}) and COBE
total root mean square CMBR temperature fluctuation measurement
(\ref{cobedt}) gives an isocurvature skewness (\ref{sapp}) of
\beq
S_{3,{\rm iso}} \sim -1 \times 10^{30}
\left( { f_a/N \over 10^{16}~{\rm GeV} } \right)^{7/2}
~ \sigma_{\theta}^4 \left( 3 \theta_i^2 + \sigma_{\theta}^2 \right)
\label{sisolowf}
\eq
For the case of larger Peccei-Quinn scales
in which axion condensate forms during a radiation
dominated era but in effectively the zero temperature axion potential and
again without any dilution after its formation, the
relic axion density is given by (\ref{relicdensitylargef})
in section 2.1.
This relic density with $f_c \xi(T_{\rm osc})^{-1} \simeq 1$ and ignoring
anharmonic effects, along with
(\ref{wmapomegam}) and (\ref{cobedt})
gives an isocurvature skewness (\ref{sapp}) of
\beq
S_{3,{\rm iso}} \simeq - 1.9 \times 10^{28}
\left( { f_a/N \over 10^{16}~{\rm GeV} } \right)^{9/2}
~ \sigma_{\theta}^4 \left( 3 \theta_i^2 + \sigma_{\theta}^2 \right)
\label{sisohighf}
\eq
Finally, for the case in which the axion condensate forms during
a matter dominated era with subsequent
late decay of the dominating particle with a reheat temperature
$T_{\rm RH} \lsim \lqcd$, the relic density is given by (\ref{rhbound}) in
section 2.3.1.
This relic density with $f_c \xi(T_{\rm osc})^{-1} \simeq 1$ and ignoring
anharmonic effects, along with
(\ref{wmapomegam}) and (\ref{cobedt})
gives an isocurvature skewness (\ref{sapp}) of
\beq
S_{3,{\rm iso}} \simeq -6.7 \times 10^{21}
\left( { T_{\rm RH} \over 6~{\rm MeV} } \right)^3
\left( { f_a/N \over 10^{16}~{\rm GeV} } \right)^{6}
~ \sigma_{\theta}^4 \left( 3 \theta_i^2 + \sigma_{\theta}^2 \right)
\label{sisotrf}
\eq
Note that the axion fluctuation
induced isocurvature skewness
is negative definite and strictly bounded by
$S_{3,{\rm iso}} \geq -\sqrt{8}$,
and as discussed below is already bounded to be significantly smaller
in magnitude.
So for a string/M-theory axion with $f_a/N \sim 10^{16}$ GeV,
similar to the case of isocurvature power fraction,
the large prefactors in (\ref{sisolowf})--(\ref{sisotrf})
imply that
even just the magnitude of the observed temperature fluctuations
(\ref{cobedt}) places a stringent limit on the combination
$\sigma_{\theta}^4 ( 3 \theta_i^2 + \sigma_{\theta}^2 )$
of variance and average value of the initial misalignment angle
in any cosmological scenario for the formation of the axion
condensate in which the axion exists during inflation.

The dimensionless skewness (\ref{eqn:skewness})
may be measured or bounded from the observed CMBR
temperature fluctuation three point function.
As reviewed in appendix C, it has become common
in the literature to parameterize
possible non-Gaussianity in terms of a non-linear product
of Gaussian-distributed temperature fluctuations proportional
to a dimensionless parameter $f_{NL}$.
As derived in (\ref{sfnl}) of appendix C, the
dimensionless skewness may be related to this parameter at leading order by
\begin{eqnarray}
S_3 \simeq 18 f_{NL} %\langle ( \delta T / T)^2 \rangle^{1/2}
\av{( \delta T / T)^2_{\rm tot} }^{1/2}
\end{eqnarray}
The WMAP bound on $f_{NL}$ for multipoles
$\ell < 65$ is roughly $f_{NL} =-150^{+200}_{-150}$ \cite{Komatsu:2003fd}.
Since the dimensionless skewness for axion induced non-Gaussianity
is strictly negative, $S_{3,{\rm iso}} \leq 0$,
we conservatively take $f_{NL} > -300$ for this case.
With the total root mean square CMBR temperature fluctuation (\ref{cobedt})
measured by COBE \cite{COBE} this bound then corresponds
to $S_{3,{\rm iso}} \gsim -6 \times 10^{-2}$.

Since the magnitude of isocurvature temperature fluctuations
falls rapidly beyond the Sachs-Wolfe plateau, as discussed in section
3.2, the inclusion of multipoles much beyond $\ell \gsim 65$ would
not significantly improve the search for, or bound on, a non-Gaussian component
in isocurvature temperature fluctuations.
And since current measurements at these low multipoles are already
cosmic variance limited, future measurements will also not significantly
improve the current bound on a non-Gaussian component in
isocurvature temperature fluctuations (although
the bound on a non-Gaussian component in adiabatic temperature fluctuations,
which do not fall as rapidly at large multipoles, would be improved).
In addition, future reductions in degeneracies of parameters
describing the dominant adiabatic power spectrum from measurements
at higher multipoles will also not help improve the bound
on a non-Gaussian component in isocurvature temperature fluctuations.
So unlike the fractional isocurvature power, $\alpha$, discussed
in section 3.2, the bound on the isocurvature skewness,
$S_{3,{\rm iso}}$ is unlikely to improve significantly.

%Above the Sachs-Wolfe plateau, the baryon--photon oscillations induced
%by the axion perturbation must be taken into account, and the analysis
%becomes substantially more complicated.  As we will see, adding this
%analysis would not qualitatively affect our conclusions; so,

% --------------------------------------------------------
% --------------------------------------------------------

\section{Constraints on the Axion from the CMBR}
\label{sec:Combine}

It seems unlikely that a
string/M-theory QCD axion with $f_a/N \sim 10^{16}$ GeV
will ever be probed directly in laboratory experiments.
Such an axion can, however, give rise to in principle measurable
isocurvature \cite{AxionIso} and non-Gaussian \cite{lythstewart}
signatures in CMBR temperature fluctuations,
as reviewed in section 3.
In this section we show that a detection of primordial gravity waves
interpreted as arising from an early epoch of inflation
places a lower limit on these observable effects.
%isocurvature and
%non-Gaussian effects of a QCD axion in CMBR temperature fluctuations.
The current bounds on isocurvature and non-Gaussian components
of CMBR temperature fluctuations are then shown to be sufficient
to rule out a string/M-theory QCD axion if gravity waves
interpreted as arising from inflation with a Hubble constant
$H_{\rm ing} \gsim 10^{13}$ GeV
are observed %by the PLANCK polarimetry experiment
by future experiments.

Consider first the axion fluctuation induced signature of an
isocurvature component in CMBR temperature fluctuations.
The goal here is to assess how well CMBR experiments
can probe a string/M-theory QCD axion independent
of any cosmological assumptions.
This requires considering in turn all the relevant cosmological
scenarios for formation of the axion condensate presented
in section 2.
In addition it requires considering the smallest possible
axion induced isocurvature component in CMBR temperature
fluctuations consistent with other future CMBR measurements;
this implies the weakest possible bound.
The numerical relations between the fractional isocurvature
power $\alpha$ and variance and average value of the initial
misalignment angle
for the various cosmological scenarios for formation
of the axion condensate considered in section 2 are given in
(\ref{alphalowf})--(\ref{alphatrf}) of section 3.2.
In each case the fractional isocurvature power is proportional
to the combination
$\sigma_{\theta}^2 ( 2 \theta_i^2 + \sigma_{\theta}^2 )$.
This is clearly minimized in the limit in which the
average misalignment angle is much smaller than the
variance of inflationary fluctuations in the initial
misalignment angle, $\theta_i^2 \ll \sigma_{\theta}^2$.
In this limit of small average misalignment angle,
the relic axion density is due mostly to inflationary
fluctuations of the axion field as discussed in section 2.4.
And the fractional axion fluctuations are order one
in this limit,
$\langle S_a^2 \rangle  \simeq 2$ from (\ref{Sthetarel}).
%are $\chi^2$ distributed as discussed section 3.3.

Axion induced isocurvature components in
CMBR temperature fluctuations of course require
primordial fluctuations in the axion field.
%As discussed in section 2.4.
As long as the Peccei-Quinn symmetry
remains unbroken during inflation, a
string/M-theory QCD axion undergoes quantum fluctuations
during this epoch with magnitude (\ref{sigmaHf}) determined by
the inflationary Hubble constant.
As discussed in section 3.1 all massless fields undergo
quantum fluctuations during inflation including the metric
which ultimately gives rise to tensor B-modes in CMBR
temperature fluctuations.
Observation of these modes interpreted as arising from
inflation would establish the Hubble constant during inflation and the
variance of fluctuations in the axion initial misalignment
angle for a given Peccei-Quinn scale, as given in
(\ref{sigmarrel}) in terms of the tensor to scalar ratio.
This would in turn establish a lower limit
for the isocurvature power fraction.
So obtaining a definitive lower limit on an axion induced
isocurvature component in CMBR temperature fluctuations
independent of cosmological assumptions requires the observation of
primordial
gravity waves %contributions to CMBR temperature fluctuations
interpreted as arising from inflation.

The lower limit on $\alpha$ in terms of the Hubble constant during inflation
for a given cosmological
scenario for the formation of the axion condensate
may be obtained from (\ref{alphalowf})--(\ref{alphatrf})
in the limit $\theta_i^2 \ll \sigma_{\theta}^2$ discussed above,
along with the relation (\ref{sigmarrel}) between the variance of the axion
initial misalignment angle and inflationary Hubble constant.
This may also be obtained directly from (\ref{alphasa})
in the limit $\langle S_a^2 \rangle  \simeq 2$ discussed
above, along with the minimum fluctuation dominated relic axion densities
(\ref{eqn:firstreldensity})--(\ref{eqn:lastreldensity})
in terms of the inflationary Hubble constant.
If gravity wave contributions to CMBR temperature fluctuations
are in fact observed and interpreted as arising from inflation,
then the %relations (\ref{alphaboundlowf}) and (\ref{alphaboundhighf})
lower limit on $\alpha$
can be inverted to give a minimum allowed Peccei-Quinn scale
in terms of the implied Hubble constant during inflation and
the present bound on the isocurvature power fraction.

%%%%%%%%%%%%%%%%%%%%%%%%%%%%%%%%%%%%%%%%%%%%%%%%%%%%%%%%%%%%5
%
\sfig{isobound}{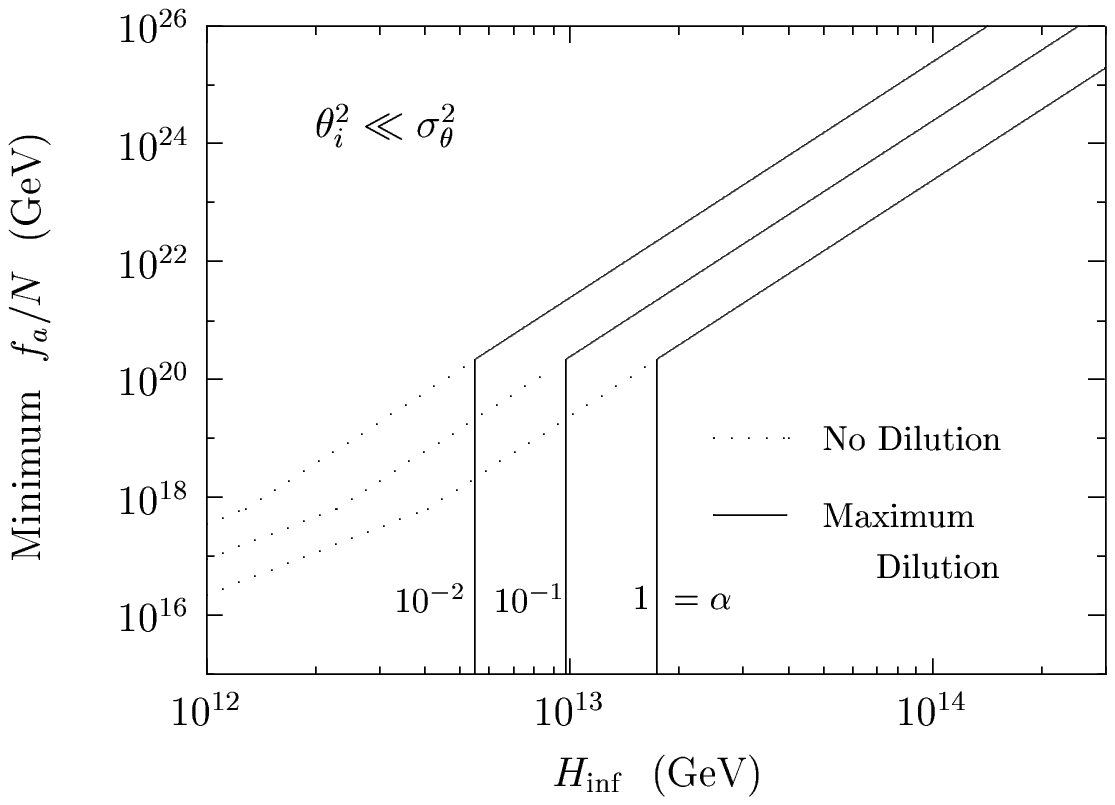}{169}{342}{482}{555}{4.75in}{0.25in}{
Minimum allowed axion Peccei-Quinn scale $f_a/N$ as a function of
$\Hinf$ for isocurvature temperature fluctuation
power fractions $\alpha=\langle (\delta T/T)^2_{\rm iso} \rangle /
\langle (\delta T/T)^2 \rangle =
1,10^{-1},10^{-2}$ in the
limit that the average initial misalignment angle is small compared
to inflation induced fluctuations,
$\theta_i^2 \ll \sigma_{\theta}^2=(\Hinf/(2 \pi f_a/N))^2$.
For $\theta_i^2 \gsim \sigma_{\theta}^2$ the minimum allowed
$f_a/N$ grows like  %$(\theta_i/ \sigma_{\theta})^2$.
$\theta_i^2$.
The dotted lines correspond to no dilution of the axion
condensate after formation while the solid lines correspond
to maximum possible dilution by a late decaying particle
with $T_{\rm RH} \simeq 6$ MeV.}
%
%%%%%%%%%%%%%%%%%%%%%%%%%%%%%%%%%%%%%%%%%%%%%%%%%%%%%%%%%%%%%5

The simplest cosmological scenario %for formation the axion condensate
to consider is obtained if the axion condensate is formed during the radiation
dominated era without any dilution after its formation.
In this case for small values of Peccei-Quinn scale
the isocurvature power fraction (\ref{alphalowf})
with $\theta_i^2 \ll \sigma_{\theta}^2$ and
the relation (\ref{sigmarrel}) yields a lower limit on the
Peccei-Quinn scale as outlined above in terms of $\alpha$
and the Hubble constant during inflation of
%\be
% \alpha
%\gsim 4 \times 10^4 \left(\frac{H_{\rm inf}}{10^{13}~{\rm
%    GeV}}\right)^4\left(\frac{10^{16}~{\rm GeV}}{f_a/N}\right)^{5/3}
%       ~~~~~~~~\theta_i^2 \lsim \sigma_{\theta}^2
%\label{alphaboundlowf}
%\ee
\be
f_a/N
\gsim {  6 \times 10^{18} \over \alpha^{3/5} } ~
 \left(\frac{H_{\rm inf}}{10^{13}~{\rm GeV}}\right)^{12/5}
%       \left(\frac{10^{16}~{\rm GeV}}{f_a/N}\right)^{5/3}
~{\rm GeV}
        ~~~~~~~~\theta_i^2 \lsim \sigma_{\theta}^2
\label{alphaboundlowf}
\ee
In the same case for large values of Peccei-Quinn scale,
the isocurvature power fraction (\ref{alphahighf})
with $\theta_i^2 \ll \sigma_{\theta}^2$ and
the relation (\ref{sigmarrel})
yields a lower limit on the Peccei-Quinn scale of
%$\alpha$ in terms
%of the Hubble constant during inflation and Peccei-Quinn scale of
%\be
%\alpha \gsim 2.4\times 10^3 \left(\frac{H_{\rm inf}}{10^{13}~{\rm
%    GeV}}\right)^4\left(\frac{10^{16}~{\rm GeV}}{f_a/N}\right)
%       ~~~~~~~~\theta_i^2 \lsim \sigma_{\theta}^2
%\label{alphaboundhighf}
%\ee
\be
f_a/N \gsim {2.4 \times 10^{19} \over \alpha} ~
     \left(\frac{H_{\rm inf}}{10^{13}~{\rm GeV}}\right)^4
%       \left(\frac{10^{16}~{\rm GeV}}{f_a/N}\right)
   ~{\rm GeV}
        ~~~~~~~~\theta_i^2 \lsim \sigma_{\theta}^2
\label{alphaboundhighf}
\ee
%If gravity wave contributions to CMBR temperature fluctuations
%are in fact observed and interpreted as arising from inflation,
%then the relations (\ref{alphaboundlowf}) and (\ref{alphaboundhighf})
%can be inverted to give a minimum allowed Peccei-Quinn scale
%in terms of the implied Hubble constant during inflation and
%the present bound on the isocurvature power fraction.
These bounds on the Peccei-Quinn scale are plotted in Fig. 2
as the diagonal dotted and solid lines for $\alpha=1,10^{-1}, 10^{-2}$.
The bounds for large and small Peccei-Quinn scale coincide for
$f_a/N \sim 6 \times 10^{17}$ GeV.
As discussed in section 2.1 for Peccei-Quinn scales
in this transition region the relic axion density calculation suffers
unknown strong QCD uncertainties but should asymptotically approach
the limiting expressions for scales well outside the transition region.

In the remaining cosmological scenario to consider, the axion condensate is
formed during a matter dominated era with subsequent decay of
the dominating particle.
In this case
with $\theta_i^2 \ll \sigma_{\theta}^2$,
the maximum possible dilution consistent with BBN
discussed in section 2.3.1, $T_{\rm RH} \simeq 6$ MeV, and
the relation (\ref{sigmarrel}) between the variance of the
initial misalignment angle and Hubble constant during inflation,
the isocurvature power fraction (\ref{alphatrf}) is bounded by
\beq
\alpha \lsim 1.1 \times 10^{-1} ~
  \left( { T_{\rm RH} \over 6~{\rm MeV} } \right)
 \left(\frac{H_{\rm inf}}{10^{13}~{\rm GeV}}\right)^4
 ~~~~~~~~\theta_i^2 \lsim \sigma_{\theta}^2
 \label{alphaboundtrh}
\eq
Note that this bound is {\it independent} of the Peccei-Quinn
scale since the explicit $(f_a/N)^4$ dependence in (\ref{alphatrf})
cancels the $(f_a/N)^{-4}$ dependence in $\sigma_{\theta}^4$.
This can also be seen directly in (\ref{alphasa})
with $\langle S_a \rangle \simeq 2$ as discussed above in this
limit, along with the minimum inflationary
fluctuation induced relic axion density (\ref{eqn:lastreldensity})
which is independent of Peccei-Quinn scale as discussed in section 2.4.
This implies that if the bound (\ref{alphaboundtrh}) is not
satisfied, then for an axion which exists during inflation,
all Peccei-Quinn scales are excluded for which the axion condensate
forms during the matter dominated era before subsequent decay of
the dominating particle.
As discussed in section 2.4, this occurs with maximum dilution,
$T_{\rm RH} \sim 6$ MeV, for Peccei-Quinn scales
$f_a/N \lsim 2 \times 10^{20}$ GeV.
For larger Peccei-Quinn scales the axion condensate forms
during the subsequent radiation dominated era after reheating from
decay of the dominating particle.
So for $f_a/N \gsim 2 \times 10^{20}$ GeV the bound (\ref{alphaboundhighf})
is recovered in any scenario for formation of the axion
condensate.
The bounds obtained from (\ref{alphaboundtrh}) for
$\alpha=1,10^{-1},10^{-2}$ with maximum dilution,
$T_{\rm RH} \sim 6$ MeV, are
plotted as vertical solid lines in Fig. 2.
Since the axion condensate forms in essentially the zero temperature
potential in this case, these bounds do not suffer large QCD
uncertainties.

Now consider the axion fluctuation induced signature of a
non-Gaussian component in CMBR temperature fluctuations.
The numerical relations between the axion induced isocurvature
skewness, $S_{3,{\rm iso}}$, and variance and average value of the
initial misalignment angle for the various cosmological scenarios
for formation of the axion condensate considered in section 2
are given in (\ref{sisolowf})--(\ref{sisotrf}) of section 3.3.
In each case the skewness is proportional to the combination
$\sigma_{\theta}^4(3 \theta_i^2 + \sigma_{\theta}^2)$.
Just as for the isocurvature power fraction,
this combination is minimized for
$\theta_i^2 \ll \sigma_{\theta}^2$ which yields in turn the
weakest possible bound.
In this limit of small initial misalignment angle, the relic
axion density is due mostly to inflationary fluctuations of
the axion field as discussed in section 2.4.
And the axion induced isocurvature temperature fluctuations
are approximately $\chi^2$ distributed as discussed in section 3.3.
The lower limit on $S_{3,{\rm iso}}$ may be obtained in the
$\theta_i^2 \ll \sigma_{\theta}^2$ limit from
(\ref{sisolowf})--(\ref{sisotrf}) along with the relation
(\ref{sigmarrel}) between
the variance of the axion initial misalignment angle and inflationary
Hubble scale.
This may also be obtained directly from (\ref{sapp})
in this limit along with the minimum fluctuation dominated
relic axion densities
(\ref{eqn:firstreldensity})--(\ref{eqn:lastreldensity})
in terms of the inflationary Hubble constant.
Just as for the isocurvature power fraction,
if gravity wave contributions to CMBR temperature fluctuations
are in fact observed and interpreted as arising from inflation,
then the %relations (\ref{alphaboundlowf}) and (\ref{alphaboundhighf})
lower limit on $S_{3,{\rm iso}}$
can be inverted to give a minimum allowed Peccei-Quinn scale
in terms of the implied Hubble constant during inflation and
the present bound on the isocurvature skewness.

%%%%%%%%%%%%%%%%%%%%%%%%%%%%%%%%%%%%%%%%%%%%%%%%%%
%{169}{302}{482}{555}
\sfig{skewbound}{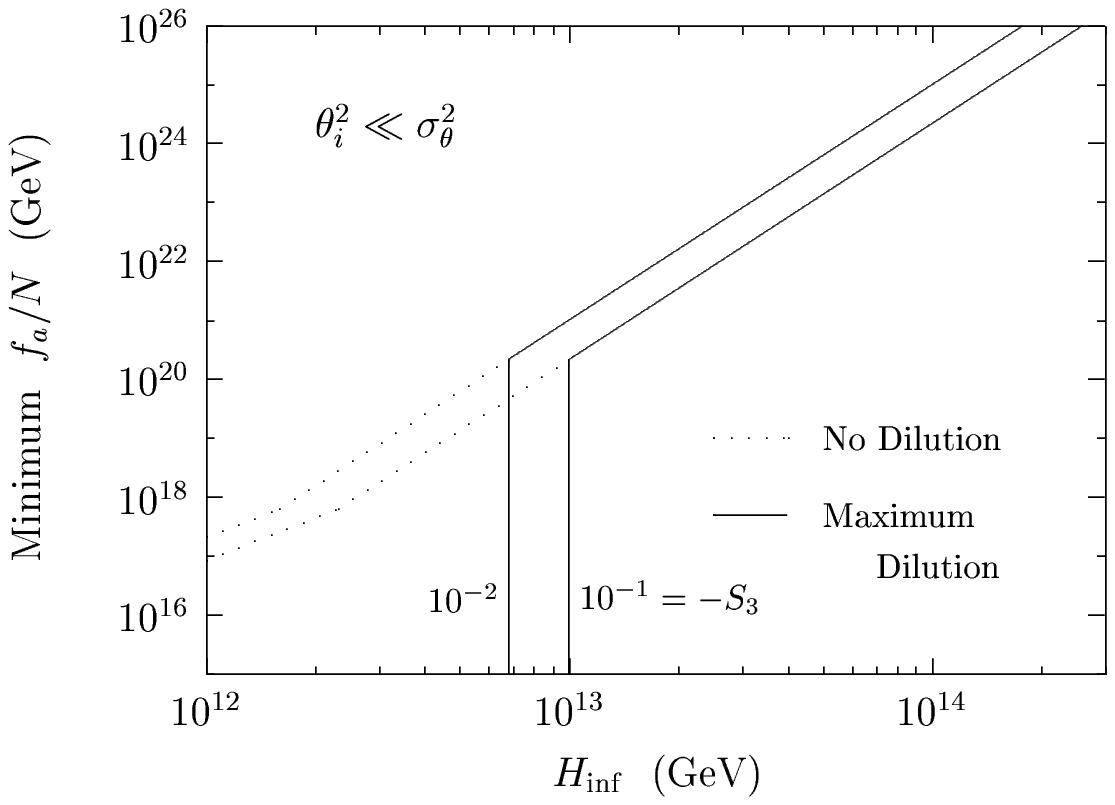}{169}{342}{482}{555}{4.75in}{0.25in}{
Minimum allowed axion Peccei-Quinn scale $f_a/N$ as a function of
$\Hinf$ for isocurvature temperature fluctuation
skewness $S_3=\langle (\delta T/T)^3_{\rm iso} \rangle /
\langle (\delta T/T)^2 \rangle^{3/2} = -10^{-1},-10^{-2}$ in the
limit that the average initial misalignment angle is small compared
with inflation induced fluctuations,
$\theta_i^2 \ll \sigma_{\theta}^2=(\Hinf/(2 \pi f_a/N))^2$.
For $\theta_i^2 \gsim \sigma_{\theta}^2$ the minimum allowed
$f_a/N$ grows like  %$(\theta_i/ \sigma_{\theta})^{4/3}$.
$\theta_i^{4/3}$.
The dotted lines correspond to no dilution of the axion
condensate after formation while the solid lines correspond
to maximum possible dilution by a late decaying particle
with $T_{\rm RH} \simeq 6$ MeV.}
%
%%%%%%%%%%%%%%%%%%%%%%%%%%%%%%%%%%%%%%%%%%%%%%%%%%

The simplest cosmological scenario for formation of the axion condensate
is that it formed during the radiation dominated era without
any dilution after its formation.
In this case for small values of Peccei-Quinn scale the isocurvature
skewness (\ref{sisolowf}) yields a lower limit of the
Peccei-Quinn scale in terms of $S_{3,{\rm iso}}$
and the Hubble constant during inflation, as described above, of
\be
f_a/N
\gsim {  8 \times 10^{18} \over (-S_{3,{\rm iso}})^{2/5} } ~
 \left(\frac{H_{\rm inf}}{10^{13}~{\rm GeV}}\right)^{12/5}
%       \left(\frac{10^{16}~{\rm GeV}}{f_a/N}\right)^{5/3}
~{\rm GeV}
        ~~~~~~~~\theta_i^2 \lsim \sigma_{\theta}^2
\label{isoboundlowf}
\ee
In the same case for large values of Peccei-Quinn scale,
the isocurvature skewness (\ref{sisohighf})
with $\theta_i^2 \ll \sigma_{\theta}^2$ and
the relation (\ref{sigmarrel})
yields a lower limit on the Peccei-Quinn scale of
\be
f_a/N \gsim {4.8 \times 10^{19} \over (-S_{3,{\rm iso}})^{2/3}} ~
     \left(\frac{H_{\rm inf}}{10^{13}~{\rm GeV}}\right)^4
%       \left(\frac{10^{16}~{\rm GeV}}{f_a/N}\right)
   ~{\rm GeV}
        ~~~~~~~~\theta_i^2 \lsim \sigma_{\theta}^2
\label{isoboundhighf}
\ee
These bounds on the Peccei-Quinn scale are plotted in Fig. 3
as the diagonal dotted and solid lines for
$-S_{3,{\rm iso}}=10^{-1}, 10^{-2}$.
The bounds for large and small Peccei-Quinn scale coincide for
$f_a/N \sim 6 \times 10^{17}$ GeV, and suffer unknown QCD uncertainties
in the transition region.
%As discussed in section 2.1 for Peccei-Quinn scales
%in this transition region the relic axion density calculation suffers
%unknown strong QCD uncertainties but should asymptotically approach
%the limiting expression for scales well outside the transition region.

In the remaining cosmological scenario to consider, the axion condensate is
formed during a matter dominated era with subsequent decay of
the dominating particle.
In this case
with $\theta_i^2 \ll \sigma_{\theta}^2$,
the maximum possible dilution consistent with BBN
discussed in section 2.3.1, $T_{\rm RH} \simeq 6$ MeV, and
the relation (\ref{sigmarrel}) between the variance of the
initial misalignment angle and Hubble constant during inflation,
the isocurvature skewness (\ref{sisotrf}) is bounded by
\beq
-S_{3,{\rm iso}} \lsim 1.1 \times 10^{-1} ~
  \left( { T_{\rm RH} \over 6~{\rm MeV} } \right)^3
 \left(\frac{H_{\rm inf}}{10^{13}~{\rm GeV}}\right)^6
 ~~~~~~~~\theta_i^2 \lsim \sigma_{\theta}^2
 \label{isoboundtrh}
\eq
Just as for the isocurvature power fraction in this scenario
for formation of the axion condensate,
this bound is {\it independent} of the Peccei-Quinn
scale since here the explicit $(f_a/N)^6$ dependence in (\ref{sisotrf})
cancels the $(f_a/N)^{-6}$ dependence in $\sigma_{\theta}^6$.
%This can also be seen directly in (\ref{})
%with $\langle S_a \rangle \simeq 2$ as discussed above in this
%limit, along with the minimum inflationary
%fluctuation induced relic axion density (\ref{eqn:lastreldensity})
%which is independent of Peccei-Quinn scale as discussed in section 2.4.
This implies that if the bound (\ref{isoboundtrh}) is not
satisfied, then for an axion which exists during inflation,
all Peccei-Quinn scales are excluded for which the axion condensate
forms during the matter dominated era before subsequent decay of
the dominating particle.
As discussed in section 2.4, this occurs with maximum dilution,
$T_{\rm RH} \sim 6$ MeV, for Peccei-Quinn scales
$f_a/N \lsim 2 \times 10^{20}$ GeV.
For larger Peccei-Quinn scales the axion condensate forms
during the subsequent radiation dominated era after reheating from
decay of the dominating particle.
So for $f_a/N \gsim 2 \times 10^{20}$ GeV the bound (\ref{isoboundhighf})
is recovered in any scenario for formation of the axion
condensate.
The bounds obtained from (\ref{isoboundtrh}) for
$-S_{3,{\rm iso}}=10^{-1},10^{-2}$ with maximum dilution,
$T_{\rm RH} \sim 6$ MeV, are
plotted as vertical solid lines in Fig. 3.
Since the axion condensate forms in essentially the zero temperature
potential in this case, these bounds do not suffer large QCD
uncertainties.

The bounds arising from isocurvature and
non-Gaussian contributions to CMBR temperature fluctuations
presented in Figs. 2 and 3 on the Peccei-Quinn scale for an axion which
exists during inflation are quite severe if primordial
gravity waves are observed in future experiments.
As discussed in section 3.1, the search for tensor B-mode components
in CMBR temperature fluctuations with the PLANCK polarimetry
experiment may reach an ultimate sensitivity to the
Hubble constant during inflation of
$H_{\rm inf} \gsim 5 \times 10^{12}$ GeV \cite{InflationEdges,rpref},
with a likely benchmark for a positive observation of
$H_{\rm inf} \gsim 10^{13}$ GeV.
Even with the rather conservative current bound on the isocurvature
power fraction discussed in section 3.2 of
$\alpha \gsim 0.4$ \cite{GarciaBellido}, if the PLANCK experiment
did in fact observe tensor B-modes interpreted as arising
from inflation with
$H_{\rm inf} \gsim 10^{13}$ GeV,
then (\ref{alphaboundhighf})
would require that the Peccei-Quinn scale be bounded by at least
$f_a/N \gsim 6 \times 10^{19}$ GeV.
And in the same case the current bound on the isocurvature
skewness discussed in section 3.3
of $-S_3 \lsim 6 \times 10^{-2}$ \cite{Komatsu:2003fd}
with (\ref{isoboundhighf})
would require that the Peccei-Quinn scale be bounded by
at least $f_a/N \gsim 3 \times 10^{20}$ GeV.
If CMBR temperature fluctuation
tensor B-modes were discovered with magnitude not too far below
the current WMAP implied limit on the Hubble constant during
inflation discussed in section 3.1
of $H_{\rm inf} \gsim 2.8 \times 10^{14}$ GeV \cite{wmapinf}
the bounds would be even more severe.
For example, if it turned out that $H_{\rm inf} = 2 \times 10^{14}$ GeV
the current isocurvature power fraction bound would require
$f_a/N \gsim 10^{25}$ GeV, while the current isocurvature
skewness bound would require $f_a/N \gsim 5 \times 10^{25}$ GeV.

If primordial gravity waves are in fact observed,
the bound on the Peccei-Quinn scale from the present
bound on a non-Gaussian component
in CMBR temperature fluctuations would be somewhat more
stringent than that from the present bound on the isocurvature
power fraction.
However, as discussed at the end of sections 3.2 and 3.3,
the bound on isocurvature non-Gaussianity is not expected
to improve in the future, while the bound on isocurvature
power fraction should improve considerably either through
a re-analysis of current WMAP data and/or future CMBR
temperature fluctuation data at larger multipoles.
So ultimately the bound from the isocurvature power fraction
would likely be somewhat better.
In comparing the isocurvature power fraction with isocurvature
skewness bounds note that the parametric dependence
on the Hubble constant during inflation of
(\ref{alphaboundlowf}) and (\ref{isoboundlowf}) for low Peccei-Quinn
scales are identical, as are those
of (\ref{alphaboundhighf}) and (\ref{isoboundhighf}) for high Peccei-Quinn
scales.
In the case that the axion condensate is formed during
a matter dominated era,
the parametric dependence of the bounds
(\ref{alphaboundtrh}) and (\ref{isoboundtrh})
on the Hubble constant during inflation
and reheat temperature after the axion condensate forms
are different however.

For lower values of %$H_{\rm inf}$
the Hubble constant during inflation the implied bounds on
the Peccei-Quinn scale given in Figs. 2 and 3 are reduced.
But as long as $H_{\rm inf} \gsim 10^{13}$ GeV, even
with the current bounds on the CMBR isocurvature power
fraction and isocurvature skewness, the bounds imply that
in this case the only allowed scenario for formation of the
axion condensate would be during the radiation dominated
era at a temperature lower than minimum allowed reheat
temperature of $T_{\rm RH} \sim 6$ MeV
implied by consistency with BBN.
In this regime the axion condensate forms in essentially
the zero temperature potential and the relic density
calculation does not suffer any strong QCD uncertainties.
So the bounds presented here
that would apply with a positive
measurement of primordial gravity waves
by PLANCK
%polarimetry experiment which requires roughly
%$H_{\rm inf} \gsim 5 \times 10^{12}$ GeV,
are largely free from such uncertainties.

%For a Hubble constant during inflation of less than roughly
%$H_{\rm inf} \lsim 10^{13}$ GeV the bounds
%it is possible in principle
%for the axion condensate to form during a matter dominated era
%and be diluted by the subsequent decay of the dominating
%particle.
%avoid ....... definitive
%But of course violating ......
%would require a late decaying particle.

The bounds discussed above on the Peccei-Quinn
scale which would result if primordial gravity waves are observed
in a future experiment are quite severe.
However, it should be noted again that these bounds are the
weakest possible given all scenarios for formation of the axion
condensate, and average initial misalignment angle.
These weakest possible bounds are obtained in only
an extraordinarily tiny sliver of the full parameter space of
initial misalignment angles.
Outside these tiny slivers even stronger bounds would apply.
For example, even
for the minimum Hubble constant during inflation of roughly
$H_{\rm inf} \sim 10^{13}$ GeV to which the
PLANCK polarimetry experiment will be sensitive for a positive
measurement, the
bound on the Peccei-Quinn scale %from (\ref{})
quoted above of $f_a/N \gsim 3 \times 10^{20}$ GeV which would result
from the current bound on the isocurvature skewness,
corresponds to an initial misalignment angle of at most
$|\theta_i| \lsim 5 \times 10^{-9}$.
For fixed $H_{\rm inf}$ and isocurvature
skewness, the bound on the Peccei-Quinn scale
grows like $\theta_i^{4/3}$ for
$\theta_i^2 \gsim \sigma_{\theta}^2$, and for fixed
isocurvature power fraction like $\theta_i^2$.
This also implies that even if the bounds were just
barely saturated, %for $H_{\rm inf} \gsim 10^{13}$ GeV,
relic axions would comprise a very small fraction of the CDM.
This can be seen directly, for example, in the relation
(\ref{alphasa}) between the isocurvature power fraction,
fractional axion fluctuations, relic axion density,
and magnitude of the total mean square temperature fluctuations.
Since as discussed above, $\langle S_a^2 \rangle \simeq 2$ results in the
weakest possible bound,
the current bound of $\alpha \lsim 0.4$ in this limit,
along with the COBE measurement (\ref{cobedt}) of the magnitude of
temperature fluctuations, implies that if the bounds are just saturated
then $\Omega_a \lsim 1 \times 10^{-5} ~\Omega_m$ at least.
%This holds as long as the axion exists during inflation.

%It should be stressed that the bounds
%(\ref{alphaboundlowf}) and (\ref{alphaboundhighf}) are the weakest
%possible bounds on any QCD axion which exists during inflation for a given
%Hubble constant during inflation and a given bound on the
%isocurvature power fraction.

Finally,
let us consider what implications the stringent bounds
on the Peccei-Quinn scale given above, if realized by
a future measurement of primordial gravity waves, would have
on a string/M-theory QCD axion.
First, since a string/M-theory QCD axion exists
during inflation these bounds would apply.
In all cases for an observation of gravity waves which
implied a Hubble constant during inflation of
$H_{\rm inf} \gsim 10^{13}$ GeV, definitive bounds on the Peccei-Quinn
scale would result as discussed above.
For this Hubble constant during inflation, the current bound
on isocurvature power fraction would require at
least $f_a/N \gsim 6 \times 10^{19}$ GeV and
the current bound on the isocurvature skewness would require
at least $f_a/N \gsim 3 \times 10^{20}$ GeV.
These bounds would exceed the four-dimensional Planck scale
of $M_p \simeq 2.4 \times 10^{18}$ GeV.
%And those discussed above which
%would result from a positive measurement of gravity waves
%by the PLANCK polarimetry experiment would far exceed the Planck scale.
Now obtaining a Peccei-Quinn scale which is in excess of the
four-dimensional Planck scale seems problematic in a fundamental
theory of gravity.
The maximum scale for compact moduli in string/M-theory is strongly
believed to be the four-dimensional
Planck scale \cite{Banks:2003sx}; no counter example is known.
With this restriction we are led to the strong conclusion
that if primordial gravity waves interpreted
as arising from inflation are observed with an implied Hubble constant
$H_{\rm inf} \gsim 10^{13}$ GeV, which includes the range
of expected sensitivity of the PLANCK CMBR polarimetry experiment,
then a string/M-theory compact modulus can not play the
role of the QCD axion and solve the strong CP problem.

% ------------------------------------------------------
% ------------------------------------------------------

\section{Remaining Axion Windows}
\label{windows}

The bounds presented in section \ref{sec:Combine}
rather convincingly require a QCD axion which exists during
inflation to have a Peccei-Quinn scale of at least
$f_a / N \gsim 3 \times 10^{20}$ GeV
if gravity waves interpreted as arising from inflation with
Hubble constant $H_{\rm inf} \gsim 10^{13}$ GeV  are
observed by the PLANCK polarimetry experiment,
and at least $f_a / N \gsim 5 \times 10^{25}$ GeV if gravity waves are observed
just below the current WMAP bound.
As discussed above, these bounds are the weakest possible
considering all relevant cosmological scenarios for formation
of the axion condensate and all possible average
initial misalignment angles, and are therefore quite conservative.
However, it is worth discussing under what conditions these
bounds do not apply and possible loopholes for evading the bounds.

In some scenarios for realizing the Peccei-Quinn mechanism,
the QCD axion arises from spontaneous breaking of an
anomalous global $U(1)$ symmetry.
In this case the Peccei-Quinn phase transition to the broken phase could
in principle take place after inflation.
The axion field would then not exist during inflation,
so the CMBR temperature fluctuations bounds discussed
here, which depend on the existence of the axion during inflation,
would not apply.
However, in such a scenario global axion cosmic strings are
formed during the Peccei-Quinn phase transition by the Kibble
mechanism.
These cosmic strings reach a scaling distribution through axion
radiation and chopping off string loops.
The cosmic string radiated axions ultimately contribute to the relic
axion density.
The total number density
of string radiated relic axions can conservatively
be estimated to be roughly an order of magnitude more than
those arising from coherent production in the zero
mode \cite{astringref}.
And in this scenario since the QCD vacuum angle winds
around the axion strings $\theta \in (0,2 \pi N]$, the average
initial misalignment angle squared is
$\langle \theta^2 \rangle = \pi^2/3$.
So without a dilution from a late decaying particle,
the bound (\ref{newbound}) discussed in section 2.1,
including
cosmic string radiated axions, would be strengthened to roughly
$f_a/N \lsim {\rm few}~\times 10^{10}$ GeV.
With a late decaying particle with the lowest possible
reheat temperature consistent with BBN constraints,
$T_{\rm RH} \sim 6$ MeV, the bound (\ref{farhbound})
discussed in section 2.3 would
likewise be strengthened to at best
$f_a/N \lsim 10^{14}$ GeV.
The lower bound on the Peccei-Quinn scale in any scenario for
realizing the Peccei-Quinn mechanism with a QCD axion
comes from the cooling rate of supernova SN1987A
which requires roughly $f_a/N \gsim 10^{10}$ GeV \cite{raffelt}.
These upper cosmological bounds and lower astrophysical bound
define the open window for
a QCD axion in a cosmological scenario with a Peccei-Quinn
phase transition after inflation.
%This is true independent of the possibility of
%future observations of inflationary gravity waves.

In scenarios in which the QCD axion does indeed exist during inflation,
the most important consideration for applicability of the
bounds discussed in section \ref{sec:Combine} is the
observation of gravity waves either from tenor B-mode
contributions to CMBR temperature fluctuations or other means.
If these are not observed
then the allowed axion window can be quite wide with in principle
no cosmological upper bound.
In this case the Hubble constant during inflation might in fact be
unobservably small.
Searches for isocurvature or non-Gaussian components of CMBR
temperature fluctuations then do not provide definitive probes
for a QCD axion which exists during inflation.
However, without dilution of the axion condensate after formation,
exceeding the bound (\ref{newbound})
of $f_a/N \lsim 3 \times 10^{11}$ GeV
%discussed in section 2.1,
does requires a small misalignment angle.
And as discussed above this is not possible in a scenario
with a Peccei-Quinn phase transition after inflation, and so
would require that the axion exist during inflation and happen
to have a small initial misalignment angle.
Even with a late decaying particle with the lowest possible
reheat temperature consistent with BBN constraints,
$T_{\rm RH} \sim 6$ MeV, exceeding the bound (\ref{farhbound})
%discussed in section 2.3,
of $f_a/N \lsim 3 \times 10^{14}$ GeV
also requires a small misalignment angle, and
so could only be realized if the axion existed during inflation.

It is also worth considering whether there are any possible
loop holes to the bounds presented in section
\ref{sec:Combine} in the case
that gravity waves interpreted as arising from inflation
are in fact observed in future experiments.
As emphasized, application of the CMBR temperature fluctuation
isocurvature and/or non-Gaussian bounds requires the existence of a
QCD axion during inflation which undergoes quantum fluctuations,
which in turn requires that the Peccei-Quinn symmetry
remain unbroken during inflation.
It is possible in principle, however, that the Peccei-Quinn
symmetry could be broken during inflation, but restored
after inflation. %in order to allow a successful solution
%to the strong CP problem.
In this case the axion would possess a non-trivial
potential during inflation which could in principle
suppress quantum fluctuations if $m_{a,{\rm inf}} \gsim
H_{\rm inf}$, where $m_{a,{\rm inf}}$ is the mass scale of
the induced axion potential during inflation.
Such a scenario requires a significant breaking of Peccei-Quinn
symmetry during inflation.
This can be obtained in field theory
models for the origin of a QCD axion designed specifically
for this purpose \cite{Dinepqviol}.
But for large Peccei-Quinn scales the minimum of the
axion potential during inflation would have to be quite
close to the zero temperature minimum in order for the
relic axion density to be consistent with the observed
CDM density.

For a string/M-theory QCD axion it is not clear how
to obtain such a large violation of Peccei-Quinn symmetry
in the early universe, consistent with the requisite exceedingly small
violation in the current epoch for a successful
solution of the strong CP problem.
One possibility might be a spontaneous breaking of electroweak
$SU(2)_L\times U(1)_Y$ gauge symmetry during inflation
by a large expectation value along the $H_u H_d$ $D$-flat direction.
If the expectation value were sufficiently large that
all the resulting standard model quark masses where
larger than the Hubble constant, $m_q \gsim H_{\rm inf}$,
then an appropriately small dilaton expectation value during inflation
could in principle lead to gluino condensation
for $SU(3)_C$ during inflation
with $\Lambda_{\rm QCD,inf} \gg H_{\rm inf}$.
%In this scenario a potential for the phase of the gluino condensate
If in addition
$U(1)_R$ symmetry were maximally broken during inflation then
a gluino mass of at most $m_{3} \lsim H_{\rm inf}$ could be
generated \cite{DRT}.
If all these conditions were met, then
mixing of the string/M-theory QCD axion with the phase of the
gluino condensate would lead to a potential for the axion
during inflation with mass scale
$m_{a,{\rm inf}}^2 \sim m_{3} \Lambda_{\rm QCD,inf}^3 / (f_a/N)^2$.
For $\Lambda_{\rm QCD,inf}$ sufficiently large, $m_{a,{\rm inf}} \gsim
H_{\rm inf}$ might be possible which would suppress
quantum fluctuations.
This seems the only reasonable scenario for suppressing quantum fluctuations
of a string/M-theory QCD axion during inflation.
While rather contrived,
it may be kept in mind as a possible loop hole
to the bounds discussed here in the case that primordial gravity waves
are in fact observed, and considered in more detail at that time.
However, it does suffer the problem that during inflation
the sum of the phases of the Higgs condensate $H_u H_d$ and
the phase of the
$U(1)_R$ breaking gluino mass $m_3$ would have to
be quite close to the sum of these phases at zero temperature.
Otherwise the minimum of the axion potential during inflation would
not be close to the zero temperature minimum, and
would lead to a relic axion density in excess of the observed CDM density.

Another possible loophole is the form of a late entropy release which could
dilute the axion condensate after formation.
The general case of dilution by a late decaying particle
was considered in section 2.3.1, and the largest such
dilution allowed by the successful predictions of BBN was
included in the bounds presented in section 4.
An arbitrary entropy release at a temperature below that at which
the axion condensate forms,
$T_{\rm osc} \sim \lqcd$, but with reheat temperature
well in excess of $T_{\rm RH} \gsim 1$ MeV could
leave BBN unmodified while diluting the axion condensate
by an arbitrary amount.
This might in principle occur from a strongly first order
phase transition in this temperature range.
However, such a strong phase transition would require
new physics at this energy scale with significant tree-level
coupling to standard model fields in order to achieve successful
reheating.
This is certainly ruled out by the non-observation of such
physics in laboratory experiments.
An extreme version of such an entropy release would be
a very late inflation with Hubble constant $H \gsim m_a$,
but with a reheat temperature
$T_{\rm RH} \gsim 1$ MeV \cite{SavasHall}.
A reheat temperature this large with such a low Hubble
constant would again require significant tree-level coupling
of the inflaton sector at the minimum of its potential
to standard model fields at this energy scale,
and is also certainly ruled out by laboratory experiments.
Even if not already ruled out, such scenarios would have to
cope with the possible problem of over dilution or production
of the baryon asymmetry.
So an arbitrary entropy release after formation of the axion
condensate does not seem to be a credible loop hole.

% --------------------------------------------------------
% --------------------------------------------------------

\section{Conclusion}
\label{conclusion}

String/M-theory vacua generically have compact moduli which can
potentially act as the QCD axion and provide an elegant
solution of the strong CP problem.
In large classes of vacua, %string/M-theory vacua,
such as with unification of
the gauge couplings by four-dimensional renormalization group running,
the Peccei-Quinn scale for a string/M-theory QCD axion can be
large, $f_a/N \sim 10^{16}$ GeV.
Obtaining an acceptable cosmological scenario to
accommodate such an axions can be problematic.
In order for relic axion production to be consistent with
measurement of the present CDM density, a small initial misalignment
angle to minimize the relic axion density,
and possibly an late entropy release to dilute it further, are required.
In addition, the late decays of relic
saxions and axinos present problems in many
scenarios for supersymmetry breaking, which must be addressed.
But it is certainly possible to construct cosmological scenarios
with a string/M-theory QCD axion
in which all the requisite conditions are met and problems
avoided.

The goal here was to assess how well precision cosmological measurements
could probe a string/M-theory QCD axion in any cosmological
scenario which can accommodate such an axion.
This required considering all possible scenarios which could
in principle minimize the observational effects of relic axions,
consistent with possible future precision cosmological
measurements.
Establishing a minimal magnitude for observational
effects requires an independent measurement which would
imply a lower limit on the density of relic axions.
Such a measurement would be provided by the observation
of primordial gravity waves interpreted as arising from
inflation.
A measurement of the power in
primordial gravity waves would
establish the magnitude of inflationary induced quantum
fluctuations of the axion field.
For a given Peccei-Quinn scale this would in turn establish a
minimum density in the axion condensate when formed,
arising from at least the inflationary axion fluctuations.
Minimizing the relic axion density resulting from
this formation requires in turn considering a scenario with an entropy release
by a late decaying particle which dilutes the axion
condensate by the maximally allowed amount consistent with
the successful predictions of BBN.

%With this,
A measurement of the power in primordial gravity waves
would also establish the minimal contribution of relic
axion fluctuations to both isocurvature and non-Gaussian components in
CMBR temperature fluctuations.
Bounds already exist from WMAP on isocurvature and non-Gaussian
contributions to CMBR temperature fluctuations.
Given even these current bounds, we have shown that an
observation of primordial gravity
waves interpreted as arising
from an early epoch of inflation with Hubble constant
$H_{\rm inf} \gsim 10^{13}$ GeV would imply
for a QCD axion which exists during inflation, a bound on
the Peccei-Quinn scale
which is in substantial excess of the four-dimensional Planck scale.
However, the Peccei-Quinn scale for a string/M-theory compact
modulus is strongly believed to be bounded from above by the
four-dimensional Planck scale \cite{Banks:2003sx}.
The above range of power in primordial gravity waves
is expected
to be within the reach of sensitivity of the PLANCK
polarimetry experiment to tensor B-mode contributions to
CMBR temperature fluctuations.
In the future, space based gravity wave interferometers may also
reach this level of sensitivity in direct searches for background
primordial gravity waves.
All this leads to the following strong conclusion:

\vspace{-.1in}
\begin{quote}
{\it A string/{\rm M}-theory compact modulus can not play the role
of a~{\rm QCD} axion and solve the strong~{\rm CP} problem
if gravitational waves interpreted as
arising from inflation  with a Hubble constant
$H_{\rm inf} \gsim 10^{13}$ {\rm GeV} are observed in future
experiments.}
%by the {\rm PLANCK}
%polarimetry experiment.}
\end{quote}
\vspace{-.1in}

\noindent
The sensitivity to isocurvature components in CMBR temperature
fluctuations are expected to improve considerably beyond the
current bound.
%Including the possibility that non-Gaussian contribution
This implies the possibility for an interesting
corollary to the above conclusion:

\vspace{-.1in}
\begin{quote}
{\it If isocurvature and/or non-Gaussian contributions
to {\rm CMBR} temperature fluctuations are observed and interpreted
as coming from a string/M-theory {\rm QCD} axion, then
primordial gravity waves arising from an early epoch of inflation
with $H_{\rm inf} \gsim 10^{13}$ {\rm GeV} will not be observed. }
\end{quote}
\vspace{-.1in}

\noindent
These conclusions require only the assumption that the
Peccei-Quinn symmetry remains unbroken in the early universe.
With this mild assumption the bounds which imply these
conclusions are quite stringent even though they are the
weakest possible considering all relevant scenarios for the
formation of the axion condensate and initial misalignment angle.

The stringent bounds on the Peccei-Quinn scale
which would result from a positive measurement of primordial
gravity waves would
apply generally to any QCD axion model in which
the axion exists during inflation.
And the bounds would imply that any such axion with a Peccei-Quinn
scale which exceeded the implied bounds would
comprise at most a very small fraction of the CDM.

The coming generation of precision cosmological observations will provide
a wealth of data on cosmological evolution.
This data can have an direct impact on our understanding of very
early universe cosmology.
As demonstrated here, these observations can also have
a direct impact on our understanding of fundamental physics at
the highest possible energy scales.

% ----------------------------------------------------

\Acknowledgements
We would like to thank T. Banks, P. Creminelli, M. Dine, W. Fischler,
J. Garcia-Ballido, M. Kamionkowski, W. Hu, B. Metcalf, and
H. Murayama for useful discussions,
and the Aspen Center for Physics where this work was partially
completed.
The work of P.F. was supported in part by the US Department of Energy.
The work of S.T. was supported in part by the US National Science Foundation
under grant PHY02-44728.

% ---------------------------------------------------------
% ---------------------------------------------------------

\appendix

\section{The Model Independent String Axion}
\label{appmiaxion}

All perturbative string vacua posses a model-independent
axion supermultiplet
which couples universally to all gauge kinetic terms.
The relation between the
Peccei-Quinn scale, four-dimensional Planck scale, and
unified gauge coupling for this axion may be determined
from the form of the Lagrangian.
The gauge kinetic terms after compactification
to four dimensions and field redefinitions
to supersymmetric Einstein frame
are of the form \cite{witten}
\beq
\int d^2 \theta ~{1 \over 4} S ~
W^{a \alpha} W^a_{\alpha} + h.c. \supset
{1 \over 4 g^2} F^a_{\mu \nu} F^{a \mu \nu} +
{i ~\theta \over 32 \pi^2} F^a_{\mu \nu}
   \widetilde{F}^{a \mu \nu}
\label{gaugelag}
\eq
where $W^a_{\alpha}$ is the superfield strength,
$F^a_{\mu \nu}$ the field strength,
$\widetilde{F}^a_{\mu \nu} = {1 \over 2}
\epsilon_{\mu \nu \rho \sigma} F^{a \rho \sigma}$ the dual
field strength, and
\beq
S = {1 \over  g^2} + {i ~\theta \over 8 \pi^2}
\eq
the bosonic component of the axion supermultiplet.
The real part of the axion multiplet is related to the unified
value of the gauge coupling by
$$
{\rm Re}~S = { \alpha_U^{-1} \over 4 \pi }
$$
where $\alpha_U = g_U^2 / 4 \pi$.
And with the normalization (\ref{gaugelag})
the $\theta$-angle, $\theta = 8 \pi^2 {\rm Im}~S$ is periodic mod $2 \pi$.

Starting from any of the ten-dimensional string theories, the axion multiplet % four-dimensional
Lagrangian kinetic terms in supersymmetric Einstein frame
after compactification to four dimensions
are \cite{witten}
\beq
- \int d^4 \theta~ M_p^2 \ln ( S^{\dagger} + S )
\supset M_p^2 { \partial_{\mu} ( {\rm Im}S)
  \partial^{\mu} ( {\rm Im}S) \over (2 {\rm Re}S)^2 }
= { M_p^2 \over ( 8 \pi^2 )^2}
  {2 \over (\alpha_U^{-1} / 2 \pi)^2 } {1 \over 2}
  \partial_{\mu} \theta \partial^{\mu} \theta
\label{axionkinetic}
\eq
where $M_p$ is the four-dimensional reduced Planck mass,
$M_p = m_p / \sqrt{ 8 \pi}$.
In any model the vacuum angle is related to the canonically normalized
axion field and Peccei-Quinn scale by
$$
\theta = {a  \over f_a / N}
$$
where $N$ is the axion anomaly coefficient.
The axion Lagrangian kinetic term is then
\beq
{1 \over 2} \partial_{\mu} a \partial^{\mu} a =
{1 \over 2} {f_a^2 \over N^2 } \partial_{\mu} \theta
  \partial^{\mu} \theta
\label{canonicalaxion}
\eq
For the model independent string axion $N=1$ for gauge
groups realized at the first Kac-Moody level.
Comparing (\ref{axionkinetic}) and (\ref{canonicalaxion})
the Peccei-Quinn scale is then related to the four-dimensional
Planck mass and unified gauge coupling by
\beq
f_a = {\sqrt 2} ~{ \alpha_U \over 4 \pi} M_P \sim 10^{16}~
   {\rm GeV}
\eq
Previous discussions \cite{BanksDine,ChoiKim,SUSYAxions}
of the model independent axion Peccei-Quinn
scale differ from this definition
by a factor proportional to $g_U^{-2}$ because of improperly normalized
$\theta$-angle.

% ---------------------------------------------------------

\section{Zero and Finite Temperature Axion Mass}
\label{appmass}

At zero temperature the physical axion mass
results mainly from the small pion component
induced by mixing with the QCD pseudo-Goldstone
multiplet.
This may be calculated by either current algebra \cite{srednicki}
or chiral Lagrangian techniques with the result
\beq
m_a \simeq {2 \sqrt{z} \over 1+z}
  {f_{\pi} m_{\pi} \over f_a / N} \simeq 1.2 \times 10^{-9}~{\rm eV}~
  {10^{16}~{\rm GeV} \over f_a/N}
\label{axionmass}
\eq
where $z=m_u/m_d \simeq 0.56$ is the ratio of up to down current
quark masses, and $f_{\pi} \simeq 93$ MeV is the pion decay
constant.
The leading corrections to (\ref{axionmass}) are
${\cal O}(m_{\pi}^2/m_K^2)$ from Kaon mixing.

At finite temperatures above the chiral symmetry breaking
scale, $T \gsim \lqcd$, the axion mass may be determined
from the $\tqcd$-dependence of the thermal free energy.
At sufficiently high temperature, $T \gg \lqcd$,
this may be reliably
related in the dilute instanton gas approximation
to the single instanton amplitude.
In this approximation the $\tqcd$-dependent free energy
density resulting from summing over both instantons and anti-instantons
is related to the integral over instanton density $n(\rho)$ at
size $\rho$ by \cite{gpy}\footnote{Expression (5.6) of Ref. \cite{gpy} for the
$\theta$-dependence of the
free energy should contain a factor of 2 for summing over both
instantons and anti-instantons.}
\beq
F(\theta,T) \simeq - 2 \cos(\theta) ~
\int d \rho ~ n(\rho,T)
\eq
For $SU(N_c)$ gauge group with
$N_f$ light quark flavors the finite temperature
instanton density in this approximation is \cite{gpy}
\beq
n(\rho,T) = { C_{N_c} \over \rho^5}
  \left( 4 \pi^2 \over g^2(\rho) \right)^{2N_c}
  ( \chi \rho)^{N_f} ({\rm det}~m) ~e^{-8 \pi^2 / g^2(\rho)}
  e^{-f(\rho T)}
\label{instantondensity}
\eq
where $C_{N_c} = (0.260~156) \chi^{-(N_c-2)}/[(N_c-1)!(N_c-2)!]$,
$\chi=1.338~76$,
$m$ is the light quark mass matrix, and
\beq
f(x)= {\pi^2 \over 3} (2 N_c + N_f) x^2 +
  12 A(x) \left[ 1 + {1 \over 6} (N_c - N_f) \right]
\eq
where $A(x)$ is very well fit by the expression
\beq
A(x) \simeq -{1 \over 12} \ln \left(1 + { \pi^2 \over 3} x^2
    \right) +
 \alpha \left[ 1 + \gamma (\pi x)^{-3/2} \right]^{-8}
\eq
where $\alpha = 0.012~897~64$ and $\gamma=0.158~58$.
The next to leading order
two-loop renormalization group improvement of the scale dependent
gauge coupling, $g^2(\rho)$,
may be related to $\lqcd$, defined as the
pole of the inverse gauge coupling defined at this order, by
\cite{gpy}
\beq
{8 \pi^2 \over g^2(\rho) } \simeq
b \ln (1 / \rho \lqcd) +
{ b_2 \over b} \ln \ln (1 / \rho \lqcd)
\label{twoloop}
\eq
where $b={1 \over 3}(11N_c - 2N_f)$ is the coefficient of the
one-loop gauge $\beta$-function, and
$b_2={1 \over 3} [ 17 N_c^2 - N_f(13 N_c^2 - 3) / 2 N_c ]$.
At next to leading order in the instanton density
(\ref{instantondensity}) the two-loop improvement for the
scale dependence of the gauge coupling (\ref{twoloop})
must be included in the instanton amplitude $e^{-8 \pi^2/g^2(\rho)}$,
but the scale dependence of the
$4 \pi^2 / g^2(\rho)$ prefactor need only be
evaluated at the one-loop level.

With the normalizations given above, the finite temperature
axion mass squared,
\beq
F(a,T) \simeq   - m_a^2(T)(f_a/N)^2 \cos[a/(f_a/N)] =
{1 \over 2} m_a^2(T) a^2 + \cdots
\eq
in the dilute instanton gas approximation at next to leading order
is then
$$
m^2_a(T) \simeq 2 C_{N_c}
  \left( { b \over 2} \right)^{2 N_c}
  \chi^{N_f}
  {\lqcd^4 \over (f_a/N)^2}
  {  {\rm det}~m \over \lqcd^{N_f} }
  \left( \lqcd \over T \right)^{N_f+b-4} ~~~~~~~~~~~~~~
  %\times
$$
\beq
~~~~~~~~~~~~~~~~~~~~  \int dx ~x^{N_f+b-5}
  \left[
    \ln \left( { T  \over  x \lqcd } \right) \right]^{2N_c-b_2/b}
  e^{-f(x)}
\label{ftmass}
\eq
where $x= \rho T$.
In the high temperature limit
the integral is dominated by instantons of size roughly of order
$\rho \sim \pi /T$.
The integral in (\ref{ftmass}) has been evaluated numerically
for the physically relevant case of $N_c=3$ and various $N_f$
with the result that the temperature dependent
axion mass may be fit by the functional form
\cite{turner}
\beq
\xi(T) \equiv{m_a(T) \over m_a }  \simeq
  C \left( \Lambda_{\rm QCD} \over 200 ~{\rm MeV} \right)^{\alpha}
  \left( \Lambda_{\rm QCD} \over T \right)^{\beta}
  \left[ 1 - \ln\left( \Lambda_{\rm QCD} \over T \right)
    \right]^d
\label{fit}
\eeq
For the normalizations given above and correctly including
both the instanton and anti-instanton contributions with
$N_f=3$ light quark flavors relevant to the oscillations
temperatures which result for $f_a/N \sim 10^{16}$ GeV,
the numerical parameters of the fit (\ref{fit}) are
$C \simeq 0.018$,
$\alpha={1 \over 2}$, $\beta \simeq 4.0$, and
$d\simeq 1.2$ \cite{turner}.
%For $f_a/N \sim 10^{16}$ GeV the oscillation temperature
%discussed in section \ref{sec:relic}
%at which the axion condensate is formed is
For oscillation temperatures not too much larger
than $\Lambda_{\rm QCD}$ the $\ln(\Lambda_{\rm QCD}/T)$
term coming from next to leading order effects
may be neglected.
This approximation is used in deriving the oscillation temperature
given in section \ref{sec:relic}.

% ---------------------------------------------------------------

\section{Parameterizing Non-Gaussian Fluctuations}

Scalar temperature fluctuations in the CMBR may generally be
related to a gauge invariant potential function
by $\delta T/T=\Phi/3$.
One manner in which non-Gaussian temperature fluctuations
could arise is that the potential $\Phi$ is a sum
of terms which are linear and quadratic in an underlying
Gaussian distributed potential function.
As discussed in section \ref{sec:nongauss} temperature
fluctuations induced by axion fluctuations are in fact
of this form.
It has become common in the literature to parameterize
this possibility by a sum of linear and non-linear
contributions to the total potential
\be
\Phi=\Phi_L+f_{NL}\left(\Phi^2_L-\av{\Phi^2_L}\right)
\label{fnlpar}
\ee
where $\Phi_L$ is the underlying Gaussian distributed
potential function, and $f_{NL}$ is a constant
parameterizing the strength of the non-linear fluctuations
quadratic in $\Phi_L$.

This parameterization of non-Gaussian fluctuations may
be related to the dimensionless skewness (\ref{eqn:skewness})
defined in section \ref{sec:nongauss}.
To see this, first consider the mean square variance
of the temperature fluctuation with the parameterization (\ref{fnlpar})
%\be
%\av{\left(\frac{\delta T}{T}\right)^2}=
%  \frac{\sigma_{\Phi_L}^2}{3^2} \left(1 +
%  2f_{NL}^2\sigma_{\Phi_L}^2\right)
%  \simeq \frac{\sigma_{\Phi_L}^2}{9}
%\ee
\be
\av{\left( \delta T / T \right)^2}=
  \frac{\sigma_{\Phi_L}^2}{3^2} \left(1 +
  2f_{NL}^2\sigma_{\Phi_L}^2\right)
  \simeq \frac{\sigma_{\Phi_L}^2}{9}
\ee
where $\sigma^2_{\Phi_L} = \langle \Phi_L^2 \rangle$ and
the second approximate equal sign results if the
non-Gaussian non-linear contributions to the potential are
small compared with the linear Gaussian contributions,
$ f_{NL}^2 \sigma_{\Phi_L}^2 \ll 1$.
Next consider the mean cubic variance of the temperature fluctuations
with the parameterization (\ref{fnlpar}).
To leading order in $f_{NL}$ it is a product of two linear Gaussian
contributions and one non-linear quadratic Gaussian contribution to the
potential
[c.f. the leading term on the right hand side of
(\ref{dtdtdt})]
%\beq
%\av{ \left( \frac{\delta T}{T} \right) ^3 }
%\simeq {3 \over 3^3} \av{ \Phi_L \Phi_L f_{NL}
%\left(\Phi^2_L-\av{\Phi^2_L}\right) } =
%{6 \over 27} f_{NL} \sigma_{\Phi_L}^4
%\eq
\beq
\av{ \left(  \delta T/ T \right) ^3 }
\simeq {3 \over 3^3} \av{ \Phi_L \Phi_L f_{NL}
\left(\Phi^2_L-\av{\Phi^2_L}\right) } =
{6 \over 27} f_{NL} \sigma_{\Phi_L}^4
\eq
where $\langle \Phi_L^4 \rangle = 3 \sigma_{\Phi_L}^4 $.
The dimensionless skewness (\ref{eqn:skewness})
is then related to $f_{NL}$ at leading order by
%\be
%S_3= \frac{\av{\left(\frac{\delta T}{T}\right)^3}}{\av{\left(\frac{\delta
%      T}{T}\right)^2}^{3/2}} \simeq 6 f_{NL}
%          \sigma_{\Phi_L} = 18 f_{NL} \sigma_I
%       \simeq 18 F_{NL} \av{ ( \deltaT / T)^2_{\rm tot}}^{1/2}
%\ee
\be
S_3= \frac{\av{\left( \delta T / T \right)^3}}
       {\av{\left(\delta T / T \right)^2}^{3/2}} \simeq 6 f_{NL}
          \sigma_{\Phi_L} \simeq
                  %18 f_{NL} \sigma_I
         18 f_{NL} \av{ ( \delta T / T)^2_{\rm tot} }^{1/2}
         \label{sfnl}
\ee
%where %$\langle ( \delta T / T )^2_{\rm tot} \rangle \simeq \sigma_I^2$
%$\av{ ( \delta T / T )^2_{\rm tot} } \simeq \sigma_I^2$
%at this order.

% -----------------------------------------------
% -----------------------------------------------

\end{document}